\documentclass{emulateapj}
\usepackage{apjfonts}
\journalinfo{The Astrophysical Journal, 2011, in press}

\shorttitle{ALFV\'{E}N WAVE TURBULENCE}
\shortauthors{VAN BALLEGOOIJEN ET AL.}

\begin{document}

\title{Heating of the Solar Chromosphere and Corona by Alfv\'{e}n
Wave Turbulence}
\author{A. A. van Ballegooijen\altaffilmark{1},
M. Asgari-Targhi\altaffilmark{2},
S. R. Cranmer\altaffilmark{1}, and E. E. DeLuca\altaffilmark{1}}
\altaffiltext{1}{Harvard-Smithsonian Center for Astrophysics,
60 Garden Street MS-15, Cambridge, MA 02138, USA}
\altaffiltext{2}{Mathematics Department, University College London,
London, WC1E 6BT, UK}

\begin{abstract}
A three-dimensional MHD model for the propagation and dissipation of
Alfv\'{e}n waves in a coronal loop is developed. The model includes
the lower atmospheres at the two ends of the loop. The waves originate
on small spatial scales (less than 100 km) inside the kilogauss
flux elements in the photosphere. The model describes the nonlinear
interactions between Alfv\'{e}n waves using the reduced MHD
approximation. The increase of Alfv\'{e}n speed with height in the
chromosphere and transition region (TR) causes strong wave reflection,
which leads to counter-propagating waves and turbulence in the
photospheric and chromospheric parts of the flux tube. Part of the
wave energy is transmitted through the TR and produces turbulence in
the corona. We find that the hot coronal loops typically found in
active regions can be explained in terms of Alfv\'{e}n wave
turbulence, provided the small-scale footpoint motions have velocities
of 1--2 km/s and time scales of 60--200 s. The heating rate per unit
volume in the chromosphere is 2 to 3 orders of magnitude larger than
that in the corona. We construct a series of models with different
values of the model parameters, and find that the coronal heating rate
increases with coronal field strength and decreases with loop length.
We conclude that coronal loops and the underlying chromosphere may
both be heated by Alfv\'{e}nic turbulence.
\end{abstract}


\section{Introduction}
\label{sect:Intro}

It has long been assumed that the solar corona is heated by
dissipation of magnetic disturbances that propagate up from the Sun's
convection zone \citep[e.g.,][]{Alfven1947}. Convective flows
interacting with magnetic flux elements in the photosphere can produce
magneto-hydrodynamic (MHD) waves that propagate up along the flux
tubes and dissipate their energy in the corona. Also, in closed
magnetic structures the random motions of photospheric footpoints can
lead to twisting and braiding of coronal field lines, and to the
formation of thin current sheets in the corona \citep[also see][]
{Parker1972, Parker1983, Parker1994, Priest2002}. Magnetic
reconnection within such current sheets may cause impulsive heating
events, called ``nanoflares'' \citep[][]{Parker1988}. The observed
X-ray emission from solar active regions may be due to the cumulative
effects of many such coronal heating events. However, the detailed
physical processes by which the corona is heated are not yet fully
understood \citep[for reviews of coronal heating, see][]
{Aschwanden2005, Klimchuk2006}. The heating of solar active regions
has in principle two contributions: (1) energy may be injected into
the corona as a result of small-scale, random footpoint motions, or
(2) the dissipated energy may originate from a large-scale
nonpotential magnetic field such as a coronal flux rope
\citep[][]{vanB2008}. In the second case the magnetic free energy is
already stored in the corona, and does not need to be transported into
the corona from the lower atmosphere. In this paper we will focus on
the first case, i.e., we assume that the large-scale magnetic field of
the active region is close to a potential field, and that most of the
energy for coronal heating is provided by small-scale footpoint
motions.

Detailed MHD models of magnetic braiding have
been developed by many authors \citep[e.g.,][]{vanB1985, vanB1986,
Mikic1989, Berger1991, Berger1993, Longcope1994, Hendrix1996a,
Hendrix1996b, Galsgaard1996, Ng1998, Craig2005, Gudiksen2005,
Rappazzo2007, Rappazzo2008}.  This modeling has generally confirmed
Parker's ideas concerning the development of thin current layers in
magnetic fields subject to random footpoint motions. However, except
for the work by \citet{Gudiksen2005}, the above-mentioned models do
not attempt to describe the real structure and dynamics of the
photospheric magnetic field. Most models use a highly simplified
geometry in which the curvature of the coronal loops is neglected and
the background magnetic field is assumed to be uniform, ${\bf B}_0 =
B_0 \hat{\bf z}$, where $\hat{\bf z}$ is the unit vector in a
Cartesian reference frame \citep[][]{Parker1972}. The photosphere at
the two ends of the loop is represented by two parallel plates $z=0$
and $z=L$, and the magnetic field lines are assumed to be
``line-tied'' at these boundaries. The imposed horizontal flows
$(v_x,v_y)$ at the boundaries are usually taken to be {\it
incompressible}. This is very different from the converging and
diverging motions observed at the solar surface.  Observations show
that the photospheric magnetic field outside sunspots is highly
intermittent and is concentrated into small flux elements or ``flux
tubes'' with kilogauss field strengths and widths of a few $100$ km
\citep[][]{Stenflo1989, Berger2001}. These flux tubes are located in
intergranular lanes and are continually jostled about by convective
flows below the photosphere. These features of the boundary motions
are not yet accurately represented in the braiding models.

Another important aspect of the lower atmosphere is that it takes a
significant amount of time for the effects of the footpoint motions
to be transmitted into the corona. The flux tubes interact with
convective flows at the base of the photosphere, and it takes 60 to
80 seconds for an Alfv\'{e}n wave to travel from that level to the
base of the corona (a height difference of about 2 Mm). Furthermore,
Alfv\'{e}n waves are subject to strong wave reflection \citep[][]
{Ferraro1958}, as the Alfv\'{e}n speed increases from about 15 km/s
in the photospheric flux tubes to more than 1000 km/s in the corona.
On first impact with the chromosphere-corona transition region (TR),
only a small fraction of the wave energy is actually transmitted into
the corona \citep[e.g.,][]{Hollweg1981, Cranmer2005}, so it takes many
bounces for the waves to be transmitted. Therefore, the relationship
between the photospheric footpoint motions and the horizontal motions
in the low corona is very complex and is subject to significant time
delays. These delays are not included in existing models of field-line
braiding.

\begin{figure} 
\epsscale{1.15}
\plotone{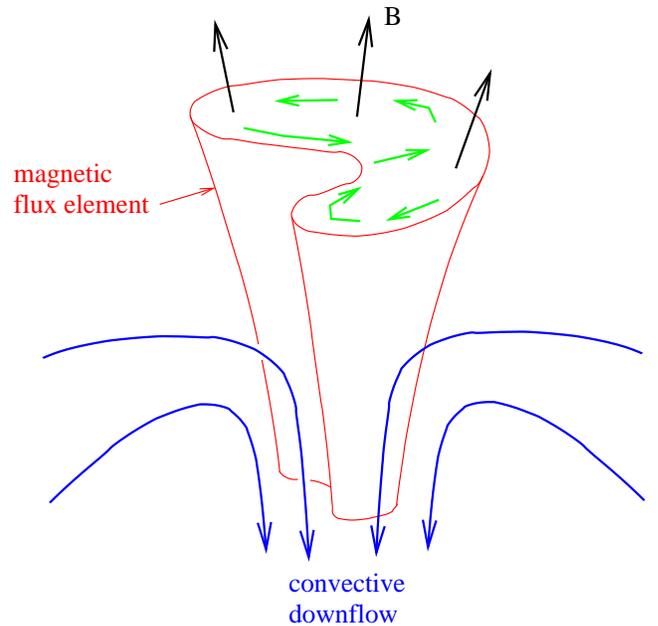}
\caption{%
Interaction of a magnetic flux element with convective flows in an
intergranular lane. The red object indicates the magnetic element
containing a nearly vertical magnetic field, as indicated by the black
arrows. The blue arrows indicate the convective flow, which push on
the flux tube from one side. Due to the stiffness of the magnetic
field, horizontal momentum is transported upward, which results in
a distortion of the shape of the flux tube and generates transverse
motions inside it (green arrows). We suggest these transverse motions
create Alfv\'{e}n waves that propagate into the upper atmosphere.}
\label{fig:cartoon}
\end{figure}

In previous models of coronal magnetic braiding it was assumed that
the photospheric footpoint motions relevant to braiding occur on a
horizontal length scale $\ell_\perp$ comparable to that of the solar
granulation or larger ($\ell_\perp >1$ Mm). Granulation flow patterns
evolve on a time scale of a few minutes, and the kilogauss flux
elements (located in the intergranular lanes) are forced to move with
these evolving flows. It was assumed that the main effect of the
granulation is to produce random {\it displacements} of the flux
elements as a whole. However, there may also exist {\it transverse
motions inside the flux elements} on a scale less than the element
width (about 100 km). This is plausible because the flux tubes are
surrounded by convective downflows that are highly turbulent
\citep[][]{Cattaneo2003, Vogler2005, Stein2006, Bushby2008}.
In the region just below the photosphere a flux element is subject to
transverse motions that not only displace the flux element as a whole,
but may also distort its shape and cause random intermixing of the
field lines inside it. The process is illustrated in Figure
\ref{fig:cartoon}, which shows a kilogauss flux tube being distorted
by convective flows that push the field lines from left to right in
the figure. At present, little is known from observations about the
magnitude of such transverse motions or the time scales involved.
However, such small-scale transverse motions could have important
effects on the upper atmosphere by producing Alfv\'{e}n waves that
propagate upward along the field lines and dissipate their energy in
the chromosphere and corona. Alfv\'{e}n waves have indeed been
observed both in the chromosphere \citep[][]{DePontieu2007b} and in
the corona \citep[e.g.,][]{Tomczyk2007}. The purpose of the present
paper is to investigate the possible role of small-scale random
motions inside photospheric flux elements in heating the solar
atmosphere.

The chromosphere is a conduit for the transport of mass and energy
into the corona. Actually, only a small fraction of the non-thermal
energy injected into the solar atmosphere is transmitted to the
corona; most of the energy is dissipated in the lower atmosphere.
Therefore, to understand the heating of the Sun's upper atmosphere it
is important to study the structure, dynamics and heating of the
chromosphere. High-resolution observations of the chromosphere have
shown that it has a complex thermodynamic structure that is strongly
influenced by the presence of magnetic fields \citep[see reviews
by][]{Judge2006, Rutten2007}. The chromosphere is highly dynamic,
and is filled with jet-like features such as mottles and dynamic
fibrils on the solar disk \citep[][]{Rouppe-van-der-Voort2007,
DePontieu2007a} and spicules
at the limb \citep[][]{DePontieu2007b}. Realistic three-dimensional
MHD models of spicule-like structures have been developed
\citep[e.g.,][]{Martinez-Sykora2009}. Internetwork regions on the
quiet Sun are affected by p-mode waves that travel upward from the
photosphere and produce shocks that cause intermittent heating
\citep[][]{Carlsson1997, Ulmschneider2005}. These shocks produce
distinct asymmetries in the profiles of the Ca~II H \& K lines.
However, the magnetic network and plage regions appear to be heated
in a different way \citep[][] {Hasan2008}. First, these regions are
{\it continually bright} in the cores of the Ca~II H \& K and Ca~II
8542 {\AA} lines, and the width of the H$\alpha$ line is enhanced
compared to the non-magnetic surroundings \citep[][]{Cauzzi2009},
indicating that the magnetic chromosphere is heated by a sustained
heating process. Second, the wavelength profiles of the Ca~II H line
from network/plage elements are {\it symmetric} with respect to the
rest wavelength \citep[e.g.,][] {Lites1993, Sheminova2005}, indicating
that the heating is not due to acoustic shock waves. In this paper we
investigate whether the heating of the magnetized chromosphere may be
due to dissipation of Alfv\'{e}n waves as suggested by
\citet{DePontieu2007b}.

The paper is organized as follows. In Section 2 the observational
constraints on braiding models are discussed. It is shown that, if the
corona is heated by dissipation of braided fields, the braiding must
occur on small transverse length scales (less than a few arcseconds).
This motivates us to develop a 3D MHD model for the dynamics of Alfv\'{e}n
waves inside a magnetic flux tube extending from the photosphere
through the chromosphere into the corona. The MHD model is described
in Section 3, and modeling results are presented in Section 4. The
results are further discussed in Section 5.

\section{Observational Constraints on Magnetic Braiding Models}

\subsection{Searching for Evidence of Braided Fields}
\label{sect:Obs}

\begin{figure}
\epsscale{1.09}
\plotone{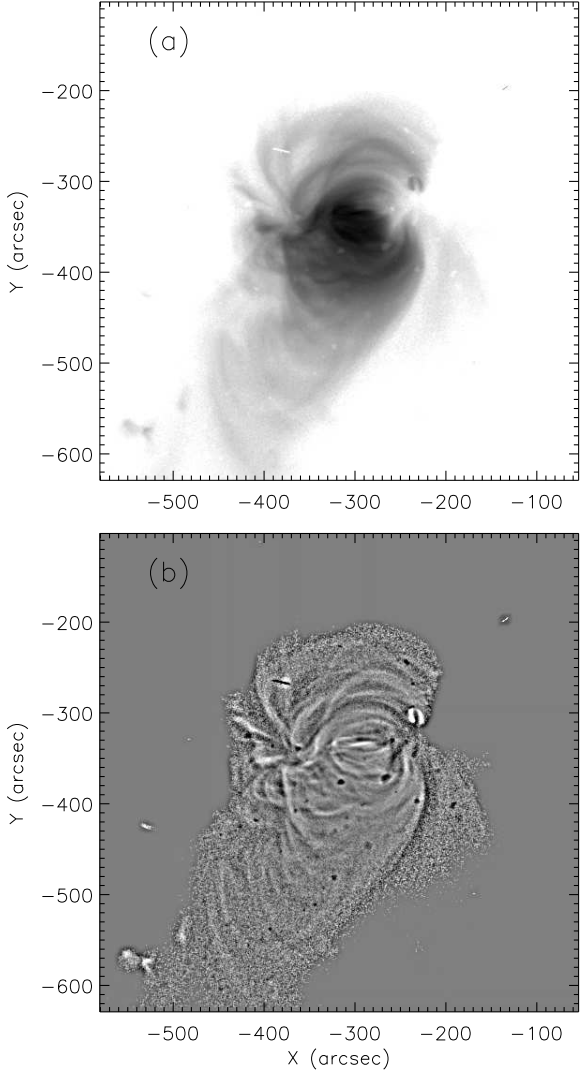}
\caption{%
(a) XRT image of active region AR 11041 on 2010 January 25, taken with
Ti-poly filter. The image is an average over the period 11:02 UT to
11:52 UT, and shows the logarithm of the intensity displayed as a
negative (black and white differ a factor 100 in intensity). The $x$
and $y$ coordinates are relative to solar disk center.
(b) Spatially filtered version of the same image (the dark spots are
artifacts due to contamination on the CCD). Note that some of the
loops appear to cross each other. The number of loop crossing is much
less than one would expect if the corona contained braided magnetic
fields on scales $\ell > 5$ arcsec.}
\label{fig:xrt1}
\end{figure}

Active regions contain loop-like structures that are aligned with the
direction of the coronal magnetic field. In this section we use data
from the X-Ray Telescope \citep[XRT,][]{Golub2007, Kano2008} on Hinode
\citep[][]{Kosugi2007} to search for braided magnetic fields in the
corona. Figure \ref{fig:xrt1}a shows active region AR 11041 observed
on 2010 January 25, starting at 11:02 UT.  The observation used the
Ti-poly filter, which is sensitive to plasma with temperatures greater
than 1 MK. In order to improve the signal-to-noise ratio, we added 12
exposures with exposure time 0.51 s, taken over a period of 50
minutes. Figure \ref{fig:xrt1}b shows a spatially filtered image where
the high spatial frequencies have been enhanced to bring out the loop
structures.

The structure of the emission is complex, and multiple structures are
superposed along the line of sight. There are some distinct coronal
loops, but there is also a more diffuse emission component. The loop
widths vary from about 3 arcsec for the narrowest loops to about 20
arcsec for the widest ones. The wider loops show some variation in
X-ray brightness across the loop on a scale of a few arcseconds, and
these fine-scale structures appear to be aligned with the loop axis.
However, our ability to observe these features is limited by the
spatial resolution of XRT and (for the outer parts of the AR) by
photon noise.

The wrapping of bright loops around each other has occasionally been
observed with the {\it Transition Region and Coronal Explorer (TRACE)}
\citep[see examples in][]{Schrijver1999}, but these cases
are often ambiguous and most loops observed with TRACE do not show
evidence for magnetic braiding. To search for braided fields with XRT,
we looked for places in Figure \ref{fig:xrt1} where two loops appear
to cross each other. There are only a few such sites within the
observed active region. If the magnetic field were braided on
observable scales (greater than 5 arcsec), one would expect to find
many more such loop crossings. The few examples that can be found seem
to involve loops that are well separated in height, and do not appear
to be due to braided structures. For the wider coronal loops, it
appears that the different threads within the loop are co-aligned to
within a few degrees, not $20^\circ$ as predicted by Parker's original
braiding model \citep{Parker1983, Priest2002, Klimchuk2006}. We
conclude there is no evidence for the existence of strongly braided
magnetic fields in the corona on spatial scales of a few arcseconds or
larger.  If there is braiding on the Sun, it must occur on small
transverse scales (less than 5 arcsec) and involve small mis-alignment
angles (at most a few degrees).

\subsection{The Coronal Heating Rate}
\label{sect:Heating}

An important constraint on any model for coronal heating is that it
must explain the average heating rate. The observed radiative and
conductive losses of active regions imply a nonradiative energy
flux into the corona $F_{mech} \sim 10^7$ $\rm erg ~ cm^{-2} ~
s^{-1}$ \citep[][]{Withbroe1977}. Assuming this energy enters a
coronal loop at both ends and is distributed uniformly over the full
length $L_{cor}$ of the loop, the average volumetric heating rate
$Q_{cor} = 2 F_{mech} / L_{cor}$. For a loop with constant cross
section and length $L_{cor} \approx 50$ Mm we obtain $Q_{cor} \sim
4 \times 10^{-3}$ $\rm erg ~ cm^{-3} ~ s^{-1}$. Another method for
estimating the heating rate is to use the scaling laws first developed
by \citet[][hereafter RTV]{Rosner1978}:
\begin{eqnarray}
T_{max} & \approx & 1.4 \times 10^3 (p_{cor} L_{cor} /2 )^{1/3}
 \nonumber \\
 & = & 1.9 \times 10^6 ~ p_{cor}^{1/3} \left( \frac{L_{cor}}
{\mbox{50 Mm}} \right)^{1/3} ~~ {\rm K} , \label{eq:Tmax} \\
Q_{cor} & \approx & 9.8 \times 10^4 p_{cor}^{7/6} (L_{cor}/2)^{-5/6}
 \nonumber \\
 & = & 1.44 \times 10^{-3} ~ p_{cor}^{7/6} \left( \frac{L_{cor}}
{\mbox{50 Mm}} \right)^{-5/6} ~ {\rm erg ~ cm^{-3} ~ s^{-1}} ,
\label{eq:QcorRTV}
\end{eqnarray}
where $T_{max}$ is the peak temperature (in K), $p_{cor}$ is the
coronal plasma pressure (in $\rm dyne ~ cm^{-2}$), and $L_{cor}$
is the loop length (in cm or Mm). X-ray observations indicate that the
loops in the core of an active region have high temperature and
pressure, $T_{max} \approx 2.5$ MK and $p_{cor} \approx 2$
$\rm dyne ~ cm^{-2}$ \citep[e.g.,][]{Saba1991, Brosius1997,
Winebarger2008, Warren2008, Ugarte-Urra2009}. Assuming a loop length
$L_{cor} \approx 50$ Mm, the required heating rate is about
$2.9 \times 10^{-3}$ $\rm erg ~ cm^{-3} ~ s^{-1}$.

Can the quasi-static braiding models explain such heating rates? In
Parker's original model \citep[][]{Parker1972, Parker1983} it is
assumed that there exist distinct ``flux tubes'' that can be traced
from the corona into the photosphere. These flux tubes are assumed to
retain their identity for about 1 day as their footpoints are moved
around on the photosphere, and thin current sheets develop at the
interfaces between the flux tubes \citep[][]{Berger1991, Berger1993,
Wilmot-Smith2009, Berger2009}. To explain the observed heating rates,
large departures from the potential field must develop. Specifically,
the angle $\theta$ between the braided field ${\bf B} ({\bf r})$ and
the potential field ${\bf B}_0 ({\bf r})$ must be about $20^\circ$
\citep[][]{Parker1983, Priest2002, Klimchuk2006}, otherwise the
Poynting flux into the corona is insufficient to balance the observed
radiative and conductive losses.  These large deviations from the
potential field are predicted to occur on transverse scales of a few
megameters or larger, and therefore should be readily observable in
the fine structures of coronal loops. As discussed in Section
\ref{sect:Obs}, there is no observational evidence for braided fields
on such scales. Therefore, it appears that Parker's original version
of the braiding model (with relatively long-lived flux tubes braided
on observable scales) is incompatible with coronal observations.

In other braiding models the boundary flows are incompressible and
vary randomly with time. This implies that any two points on the
boundary tend to separate from each other \citep[e.g.,][]{vanB1988},
hence there are no well-defined ``flux tubes'' that retain their
identity for many hours. This type of model is consistent with the
observation that photospheric flux elements frequently split up and
merge \citep[][]{Berger1996}. The magnetic fields produced in such
models are more fragmented than those predicted by Parker's original
model.  Specifically, current sheets can develop anywhere within the
volume, and the strongest currents will develop in locations where the
footpoint motions have the largest cumulative shear. Analytic studies
predict a rapid ``cascade'' of magnetic energy to small transverse
scales \citep[][]{vanB1985, vanB1986}. In fact, the cascade proceeds
so rapidly that the energy is dissipated before strong departures from
the potential field can develop. As a result, these simple cascade
models have difficulty explaining the observed rate of coronal heating
in active regions. The model predicts that the coronal heating rate
per unit volume is given by
\begin{equation}
Q_{cor} \approx \frac{B_0^2} {8 \pi} \frac{2u_0^2 \tau_0 \ln R_m}
{3L^2 \sqrt{2 \pi}} , \label{eq:Q1986}
\end{equation}
where $B_0$ is the coronal field strength, $u_0$ is the rms velocity
of the footpoint motions, $\tau_0$ is the correlation time, and $R_m$
is the magnetic Reynolds number \citep[see][]{vanB1986}. For motions
induced by the solar granulation, $\tau_0 \sim 5$ minutes, which is
slow compared to the coronal Alfv\'{e}n travel time ($\tau_0 \gg L/v_A$,
where $v_A$ is the Alfv\'{e}n speed, $v_A \sim 1000$ $\rm km ~ s^{-1}$).
Therefore, the magnetic field is expected to evolve through a series
of nearly force-free equilibrium states. Note that $Q_{cor}$ is
proportional to the product $u_0^2 \tau_0$, which is essentially the
photospheric diffusion constant, $D = \onehalf u_0^2 \tau_0
\sqrt{2 \pi}$ in this model. Therefore, measurements of $D$ can be
used as a constraint on the braiding model. Using $D \approx 250$
$\rm km^2 ~ s^{-1}$ \citep[][]{DeVore1985}, it was found that the
heating rate predicted by the cascade model is a factor 10 to 40
smaller than the heating rate observed in active regions.

Numerical simulations of magnetic braiding driven by slow, random,
incompressible footpoint motions \citep[e.g.,][]{Mikic1989,
Longcope1994, Hendrix1996a, Hendrix1996b, Galsgaard1996, Craig2005}
predict heating rates that are basically consistent with equation
(\ref{eq:Q1986}). Therefore, like the above cascade model, these
numerical models also cannot explain the observed heating rate when
realistic values for the random walk diffusion constant $D$ are
inserted into the model. We conclude that neither Parker's original
version of the braiding model nor its later modifications can
adequately explain the structure and heating of coronal loops in
active regions. This leads us to consider other types of footpoint
motion, such as small-scale random motions {\it inside} the
photospheric flux elements (see Figure \ref{fig:cartoon}).

\section{Alfv\'{e}n Wave Turbulence Model}
\label{sect:Model}

A model for the dynamics of plasma and magnetic field inside a coronal
loop is developed. We consider only a thin tube of magnetic flux,
corresponding to a single kilogauss flux tube in the photosphere.
The tube extends from the photosphere at one end of the loop, through
the chromosphere into the corona, and back down into the photosphere
at the other end. In the photosphere and chromosphere the flux tube is
assumed to be vertically oriented. We assume that a single flux tube
at one end of the loop is connected to a single flux tube at the other
end, i.e., we ignore the fact that on the real Sun the photospheric
flux concentrations at the two ends are uncorrelated and do not
perfectly match up. Also, on the real Sun magnetic flux elements
frequently split up and merge with their neighbors \citep[][]
{Berger1995, Berger1996, vanB1998}. Such processes may lead to
magnetic reconnection and may be important for coronal heating
\citep[e.g.,][]{Furusawa2000, Sakai2002}. However, in the present
model we neglect such effects, and we assume that the flux tube
retains its identity for the duration of the simulation.

The tube has a length $L$, and the overall curvature of the tube is
neglected. We use a coordinate system $(x,y,z)$, where $z$ is the
coordinate along the tube axis ($0 \le z \le L$), and $x$ and $y$ are
perpendicular to the axis. Note that near the ``left'' end of the flux
tube ($z \approx 0$) the height in the lower atmosphere is given by
$z$ and gravity acts in the $- \hat{\bf z}$ direction, but near the
``right'' end ($z \approx L$) the height is given by $L-z$ and gravity
is in the $+ \hat{\bf z}$ direction. Despite these differences, we
will sometimes refer to $z$ simply as the ``height'' in the flux tube.
The tube is assumed to have a circular cross-section with radius
$R(z)$, which is much smaller than its length $L$. To simulate the
effects of solar convection interacting with the flux tube, we impose
random footpoint motions on the field lines at the base of the
photosphere ($z=0$ and $z=L$). These motions produce Alfv\'{e}n waves
that travel along the flux tube, reflect due to gradients in
Alfv\'{e}n speed, and generate turbulence via nonlinear wave-wave
interactions. The dynamics of the waves inside the tube are described
by the equation of motion,
\begin{equation}
\rho \frac{d {\bf  v}} {dt} = - \nabla p + \rho {\bf g} +
\frac{1}{4 \pi} ( \nabla \times {\bf B} ) \times {\bf B} +
{\bf D}_v , \label{eq:dvdt} 
\end{equation}
and the magnetic induction equation,
\begin{equation}
\frac{\partial {\bf  B}} {\partial t} = \nabla \times \left(
{\bf v} \times {\bf B} \right) + {\bf D}_m . \label{eq:dBdt}
\end{equation}
Here $\rho ({\bf r},t)$ is the plasma density, $p({\bf r},t)$ is the
pressure, ${\bf v} ({\bf r},t)$ is the velocity, ${\bf B} ({\bf r},t)$
is the magnetic field, ${\bf g} = g_0 (z) \hat{\bf z}$ is the
acceleration of gravity, and ${\bf D}_v$ and ${\bf D}_m$ are viscous
and resistive dissipation terms. The magnetic field satisfies the
solenoidal condition, $\nabla \cdot {\bf B} = 0$. The velocity
${\bf v}$ can be split into parallel and perpendicular components:
\begin{equation}
{\bf v} \equiv {\bf v}_\perp + v_\parallel \hat{\bf B} ,
\label{eq:vsum}
\end{equation}
where $\hat{\bf B} ({\bf r},t)$ is the unit vector along the perturbed
magnetic field, and ${\bf v}_\perp \cdot \hat{\bf B} = 0$. Taking the
inner product of equation (\ref{eq:dvdt}) with $\hat{\bf B}
({\bf r},t)$, we obtain
\begin{equation}
\rho \frac{d v_\parallel} {dt} = \rho \left( {\bf v}_\perp \cdot
\frac{d \hat{\bf B}} {dt} \right) + \hat{\bf B} \cdot \left(
- \nabla p + \rho {\bf g} + {\bf D}_v \right) . \label{eq:vpara}
\end{equation}
The first term on the right hand side is the centrifugal force due to
changes in the shapes of the magnetic field lines, which is an
important driver of field-aligned flows. However, in the present paper
we will neglect parallel flows and focus on the perpendicular motions
of the plasma. Note that, according to equation (\ref{eq:dBdt}), the
parallel velocity $v_\parallel$ has no direct effect on the evolution
of the magnetic field.

In this paper we use the so-called reduced MHD (RMHD) approximation
\citep[][]{Strauss1976}, which assumes that the magnetic fluctuations
are small compared to the background field. In subsection
\ref{sect:RMHD} we present a derivation of the RMHD equations in the
context of the present model. In the following subsections we describe
the structure of the background atmosphere, the imposed footpoint
motions, and the numerical techniques for solving the RMHD equations.
Subsection \ref{sect:Limitations} describes some of the limitations of
the present model.

\subsection{Reduced MHD Equations for a Non-Uniform Medium With
Gravity} 
\label{sect:RMHD}

In this section we derive the RMHD equations describing the nonlinear
dynamics of Alfv\'{e}n waves in a flux tube with non-constant
cross-section and with density varying by orders of magnitude along
the flux tube. The RMHD equations were first derived by
\citet{Kadomtsev1974} and \citet{Strauss1976} for a uniform background
field \citep[also see][]{Montgomery1982, Hazeltine1983}, and the
relationships between compressible MHD, incompressible MHD and reduced
MHD were extensively discussed by \citet{Zank1992}.
\citet{Schekochihin2009} considered the extension of the RMHD
equations into the kinetic regime.
\citet{Bhattacharjee1998} included the effects of plasma pressure,
spatial inhomogeneities and parallel flows into the formalism, and
they derived a more general set of four coupled equations. However,
they did not include the effects of gravity. In the absence of
gravity, the magneto-static equilibrium equation implies that the
gradient of the plasma pressure is perpendicular to the mean magnetic
field, $\nabla p_0 \perp {\bf B}_0$. In contrast, in the present case
gravity plays an important role in the stratification of the plasma
pressure $p_0 (z)$ and density $\rho_0 (z)$ in the photosphere and
chromosphere, and at the axis of the flux tube $\nabla p_0 \parallel
{\bf B}_0$.  Therefore, the four-field equations derived by
\citet{Bhattacharjee1998} cannot be directly applied to the present
case. Also, many analyses of the RMHD equations use normalized
dynamical variables, which makes sense when the background density
is roughly constant, but not when $\rho_0$ varies by several orders
of magnitude within the system. This makes it necessary to discuss
in some detail the assumptions underlying the equations used in the
present work.

Our starting point is the equation of motion (\ref{eq:dvdt}). The
magnetic and velocity fields are written as sums over mean and
fluctuating components, ${\bf B} = {\bf B}_0 + {\bf B}_1 + \cdots$
and ${\bf v}_\perp = {\bf v}_{\perp,0} + {\bf v}_{\perp,1} + \cdots$,
and similar for the parallel velocity, density and pressure. The
background field ${\bf B}_0$ is non-uniform and varies on a spatial
scale $H_B$, which is defined by
\begin{equation}
H_B \equiv B_0 \left( \hat{\bf B}_0 \cdot \nabla B_0 \right)^{-1} ,
\label{eq:HB}
\end{equation}
where $B_0 ({\bf r})$ is the background field strength, and
$\hat{\bf B}_0 ({\bf r})$ is the unit vector along the background
field. We assume that the background atmosphere is in static
equilibrium (${\bf v}_{\perp,0} = 0$ and $v_{\parallel,0} = 0$), and
that the interior of the flux tube is current-free, $\nabla \times
{\bf B}_0 = 0$, i.e., all currents are located at the interface
between the flux tube and its surroundings. Then equation
(\ref{eq:dvdt}) yields $\nabla p_0 = \rho_0 {\bf g}$, where $p_0 (z)$
and $\rho_0 (z)$ are the unperturbed pressure and density as functions
of height $z$ inside the flux tube.

We now assume that the radius $R(z)$ of the flux tube is small
compared to the length scale $| H_B (z) |$ of the magnetic field
inside the tube. Then we can define a small parameter,
\begin{equation}
\epsilon_0 (z) \equiv \frac{R(z)} {| H_B (z) |} \ll 1 ,
\label{eq:eps0} 
\end{equation}
and the background field ${\bf B}_0 ({\bf r})$ can be approximated as
\begin{equation}
{\bf B}_0 = B_{00} \hat{\bf z} - \onehalf \frac{d B_{00}} {dz}
( x \hat{\bf x} + y \hat{\bf y} ) + {\cal O} (B_{00} \epsilon_0^2) , 
\label{eq:B0vec}
\end{equation}
where $B_{00} (z)$ is the field strength on the tube axis ($x=y=0$).
The unit vector along the background field is given by
\begin{equation}
\hat{\bf B}_0 = \hat{\bf z} - \frac{1} {2 H_B} ( x \hat{\bf x}
+ y \hat{\bf y} ) + {\cal O} (\epsilon_0^2) , \label{eq:B0hat}
\end{equation}
where we used $H_B (z) \approx B_{00} / (dB_{00}/dz)$. Flux
conservation implies that $B_{00} R^2 \approx$ constant along the flux
tube. Note that $\nabla \cdot {\bf B}_0 = 0$ as required, and that the
unit vector varies over the cross-section of the flux tube:
\begin{equation}
\frac{\partial \hat{\bf B}_0} {\partial x} = - \frac{\hat{\bf x}}
{2 H_B} + {\cal O} ( \epsilon_0^2 /R) , ~~~~~
\frac{\partial \hat{\bf B}_0} {\partial y} = - \frac{\hat{\bf y}}
{2 H_B} + {\cal O} ( \epsilon_0^2 /R) . \label{eq:dB0hat}
\end{equation}
The above ``thin tube approximation'' has been used by many authors
to study waves and instabilities in flux tubes \citep[e.g.,][]
{Defouw1976, Roberts1978, Spruit1981}.

We now consider the perturbations due to Alfv\'{e}n waves that are
launched at the base of the photosphere ($z=0$ and $z=L$) and are
reflected in the chromosphere and at the TR. The velocity amplitude
$u_\perp (z)$ of the waves is assumed to be small compared to the
Alfv\'{e}n speed $v_A (z)$, so that the magnetic perturbation
${\bf B}_1 ({\bf r}, t)$ is small compared to the background field
$B_{00}(z)$. Furthermore, we assume that the transverse length scale
$\ell_\perp$ of the waves is less than the tube radius ($\ell_\perp
\le R$) and is small compared to the parallel scale $\ell_\parallel$.
The latter is defined by
\begin{equation}
\ell_\parallel \equiv \min ( v_A \tau , | H_B | , L) ,
\end{equation}
where $\tau$ is the typical time scale of the magnetic fluctuations
(e.g., the Alfv\'{e}n wave period), $| H_B |$ is the length scale of
the background field, and $L$ is the loop length. Then we can define a
second small parameter:
\begin{equation}
\epsilon \equiv \max \left( \frac{u_\perp} {v_A} ,
\frac{\ell_\perp} {\ell_\parallel} \right) \ll 1 ,
\end{equation}
and we can expand the magnetic and velocity fields in powers of
$\epsilon$:
\begin{eqnarray}
{\bf B} & = & {\bf B}_0 + {\bf B}_1 + {\cal O} (B_{00} \epsilon^2) ,
\label{eq:Btot} \\
{\bf v}_\perp & = & {\bf v}_{\perp,1} + {\bf v}_{\perp,2} + {\cal O}
(v_A \epsilon^3) , \label{eq:vtot}
\end{eqnarray}
where in general $| {\bf B}_1 | = {\cal O} (B_{00} \epsilon)$,
$| {\bf v}_{\perp,1} | = {\cal O} (v_A \epsilon)$ and
$| {\bf v}_{\perp,2} | = {\cal O} (v_A \epsilon^2)$. We assume that
the main driver of parallel flows is the centrifugal force given by
the first term in equation (\ref{eq:vpara}). Then the parallel
velocity $v_\parallel = {\cal O} ( v_A \epsilon^2)$. 

In some derivations of the RMHD equations \citep[e.g.,][]
{Strauss1976, Montgomery1982} it is {\it assumed} that the plasma flow
is incompressible, $\nabla \cdot {\bf v} \approx 0$, but others 
have argued that this assumption cannot be taken for granted when the
plasma beta is of order unity \citep[e.g.,][]{Zank1992}. Here $\beta
\equiv 8 \pi p_0 /B_0^2$ is the ratio of gas pressure and magnetic
pressure. For the models presented in this paper, $\beta \sim 1$ in
the photospheric part of the flux tube (see section
\ref{sect:Background}). It is not clear from the literature on the
RMHD equations whether the assumption of incompressibility is still
valid when $\beta \sim 1$ and the background medium is
non-uniform. Therefore, we must first estimate the magnitude of
$\nabla \cdot {\bf v}$ for our particular model and determine whether
the assumption of incompressibility is still valid.  Using equation
(\ref{eq:vsum}), the velocity divergence can be written as
\begin{equation}
\nabla \cdot {\bf v} = \nabla \cdot {\bf v}_\perp + {\bf B} \cdot
\nabla \left( \frac{v_\parallel} {B} \right) . \label{eq:divv}
\end{equation}
The second term is of the order of $v_A \epsilon^3 / \ell_\perp$,
but the magnitude of the first term is unclear. To estimate this
term, we insert expressions (\ref{eq:Btot}) and (\ref{eq:vtot}) into
the momentum equation (\ref{eq:dvdt}), omitting the dissipative term,
and then take the divergence, which yields
\begin{equation}
\nabla^2 \left( p + \frac{B^2} {8 \pi} \right) = {\cal O}
\left( \frac{B_{00}^2} {4 \pi \ell_\perp^2} \epsilon^2 \right) .
\end{equation}
It follows that the perturbation of the total pressure is of second
order in $\epsilon$ \citep[][]{Zank1992, Bhattacharjee1998}:
\begin{equation}
p_1 + \frac{{\bf B}_0 \cdot {\bf B}_1} {4 \pi} = {\cal O}
\left( \frac{B_{00}^2} {4 \pi} \epsilon^2 \right) .
\label{eq:ptot}
\end{equation}
Also, taking the inner product of equation (\ref{eq:dBdt}) with
${\bf B}/4 \pi$, we obtain
\begin{displaymath}
\frac{\partial} {\partial t} \left( \frac{B^2} {8 \pi} \right)
+ \nabla \cdot \left( \frac{B^2} {4 \pi} {\bf v}_\perp \right)  = 
\frac{1} {4 \pi} ( {\bf v}_\perp \times {\bf B} ) \cdot ( \nabla
\times {\bf B} )
\end{displaymath}
\begin{equation}
  \approx \, - {\bf v}_\perp \cdot \left( \rho_0 \frac{d {\bf  v}}
{dt} + \nabla p_1 - \rho_1 {\bf g} \right) , \label{eq:B2}
\end{equation}
where we used equation (\ref{eq:dvdt}), again without the dissipative
term. Inserting expressions (\ref{eq:Btot}), (\ref{eq:vtot}) and
(\ref{eq:ptot}), we find
\begin{equation}
- \frac{\partial p_1} {\partial t} - {\bf v}_\perp \cdot \nabla p_1 +
\frac{B_0^2} {4 \pi} \nabla \cdot {\bf v}_\perp = {\cal O} \left(
\frac{B_{00}^2 v_A} {4 \pi \ell_\perp} \epsilon^3 \right) .
\label{eq:dpdt1}
\end{equation}
Finally, the pressure is assumed to evolve adiabatically, $dp/dt =
- \gamma p \nabla \cdot {\bf v}$, where $\gamma$ is the ratio of
specific heats. In lowest order of $\epsilon$ this equation becomes
\begin{equation}
\frac{\partial p_1} {\partial t} + {\bf v}_\perp \cdot \nabla p_1
+ \gamma p_0 \nabla \cdot {\bf v}_\perp =
{\cal O} \left( p_0 \frac{v_A}{\ell_\perp} \epsilon^3 \right) ,
\label{eq:dpdt2}
\end{equation}
where we used ${\bf v} \cdot \nabla p_0 \sim v_A \epsilon^2
| dp_0/dz |$. Adding equations (\ref{eq:dpdt1}) and (\ref{eq:dpdt2}),
and dividing by the factor $(\gamma p_0 + B_0^2 /4 \pi)$, we find
\begin{equation}
\nabla \cdot {\bf v}_\perp = {\cal O} \left( \frac{v_A}{\ell_\perp}
\epsilon^3 \right) , \label{eq:divvp1}
\end{equation}
i.e., the magnitude of $\nabla \cdot {\bf v}_\perp$ is of third
order in $\epsilon$. Inserting this result into (\ref{eq:dpdt2}),
we find $p_1 = {\cal O} ( p_0 \epsilon^2 )$, so the pressure
variations are really of second order in $\epsilon$, and we can set
$p_1 = 0$. Inserting $p_1 = 0$ into equation (\ref{eq:ptot}), we find
${\bf B}_0 \cdot {\bf B}_1 = {\cal O} (B_{00}^2 \epsilon^2 )$, so the
first-order perturbation of the magnetic field is perpendicular to the
background field, ${\bf B}_1 \perp {\bf B}_0$. Therefore, the effects
of compressibility can be neglected in the present models, $\nabla
\cdot {\bf v}_\perp = 0$, even though $\beta \sim 1$ in some regions
of the model. This is due to the fact that the transverse motions of
the waves are nearly horizontal and therefore along the planes of
constant pressure $p_0 (z)$ inside the flux tube.

The induction equation (\ref{eq:dBdt}) can also be written as
\begin{equation}
\frac{\partial {\bf A}} {\partial t} = {\bf v}_\perp \times {\bf B}
+ \nabla \phi , \label{eq:dAdt}
\end{equation}
where ${\bf A} ({\bf r},t)$ is the vector potential (${\bf B} \equiv
\nabla \times {\bf A}$), $\phi ({\bf r},t)$ is the electric potential,
and we omit the dissipative term. From the above analysis it is clear
that the Lorentz force is of order $\epsilon^2$, so the component of
electric current perpendicular to ${\bf B}_0$ is also of this order
\citep[][]{Strauss1997}. Therefore, the first-order perturbation of
the vector potential must be parallel to the background field:
\begin{equation}
{\bf A}_1 ({\bf r},t) = h({\bf r},t) {\bf B}_0 ({\bf r}) ,
\label{eq:A1}
\end{equation}
where $h$ is the magnetic flux function ($h \sim \ell_\perp
\epsilon$). Using $\nabla \times {\bf B}_0 = 0$, we find ${\bf B}_1
= \nabla h \times {\bf B}_0$, which is perpendicular to ${\bf B}_0$ as
required. Then the total magnetic field is
\begin{equation}
{\bf B} ({\bf r},t) = {\bf B}_0 + \nabla_\perp h \times {\bf B}_0
+ {\cal O} ( B_{00}^2 \epsilon^2 ) . \label{eq:B1}
\end{equation}
Here $\nabla_\perp$ is defined in relation to the background field,
$\nabla_\perp \equiv \nabla - \hat{\bf B}_0 ( \hat{\bf B}_0 \cdot
\nabla)$, whereas ${\bf v}_\perp$ is defined to be perpendicular to
the perturbed field ${\bf B}$. Inserting expression (\ref{eq:A1}) into
(\ref{eq:dAdt}) and taking the inner product with ${\bf B}$, we
obtain
\begin{equation}
\frac{\partial h} {\partial t} = \frac{1}{B_0^2} {\bf B} \cdot \nabla
\phi = \frac{1}{B_0^2} {\bf B}_0 \cdot \left( \nabla \phi
+ \nabla_\perp \phi \times \nabla_\perp h \right) , \label{eq:dhdt1}
\end{equation}
where we use ${\bf B} \cdot {\bf B}_0 = B_0^2$. Taking the cross
product of (\ref{eq:dAdt}) with ${\bf B}_0$, and using the condition
${\bf v}_\perp \cdot {\bf B} = 0$, we find
\begin{equation}
{\bf v}_\perp ({\bf r},t) = \frac{1}{B_0^2} \left( \nabla_\perp \phi
\times {\bf B}_0 + \xi {\bf B}_0 \right) + {\cal O} ( v_A \epsilon^3 )
, \label{eq:vperp1}
\end{equation}
where $\xi \equiv - \nabla_\perp \phi \cdot \nabla_\perp h$. The two
terms with $\phi$ and $\xi$ are of order $v_A \epsilon$ and $v_A
\epsilon^2$, respectively. Taking the divergence of equation
(\ref{eq:vperp1}), we obtain
\begin{equation}
\nabla \cdot {\bf v}_\perp = \left( \nabla_\perp \phi
\times {\bf B}_0 + \xi {\bf B}_0 \right) \cdot \nabla \left(
\frac{1}{B_0^2} \right) + \frac{1}{B_0^2} {\bf B}_0 \cdot \nabla \xi
+ {\cal O} \left( \frac{v_A}{\ell_\perp} \epsilon^3 \right) .
\label{eq:divvp2}
\end{equation}
Estimating the magnitudes of the various terms, and using the fact
that $\nabla B_0$ is nearly parallel to ${\bf B}_0$, we find that all
terms are of the order of $v_A \epsilon^3 / \ell_\perp$, so equation
(\ref{eq:divvp2}) is consistent with equation (\ref{eq:divvp1}).
Neglecting terms of order $\epsilon^2$ in equation (\ref{eq:vperp1}),
the perpendicular velocity can be further approximated as
\begin{equation}
{\bf v}_\perp = \nabla_\perp f \times \hat{\bf B}_0
+ {\cal O} (v_A \epsilon^2) , \label{eq:vperp2}
\end{equation}
where $f \equiv \phi ({\bf r},t) / B_0 ({\bf r})$. In this
approximation the component of ${\bf v}_\perp$ parallel to ${\bf B}_0$
is neglected. The quantity $f({\bf r},t)$ can be interpreted as the
velocity stream function. Inserting the expression $\phi = f B_0$ into
equation (\ref{eq:dhdt1}), we obtain for the magnetic induction
equation: 
\begin{equation}
\frac{\partial h} {\partial t} = \hat{\bf B}_0 \cdot \nabla f
+ \frac{f} {H_B} + \hat{\bf B}_0 \cdot ( \nabla_\perp f \times
\nabla_\perp h ) . \label{eq:dhdt2}
\end{equation}
We now consider the equation of motion (\ref{eq:dvdt}). As discussed
above, we can neglect the fluctuations in gas pressure and density
($p_1 = \rho_1 = 0$), and we also omit the viscous term. Using
equations (\ref{eq:dB0hat}) and (\ref{eq:B1}), the curl of the
magnetic field can be approximated as
\begin{eqnarray}
\nabla \times {\bf B} & = & {\bf B}_0 \cdot \nabla ( \nabla_\perp h )
- \nabla_\perp h \cdot \nabla {\bf B}_0
- [ \nabla \cdot ( \nabla_\perp h ) ] {\bf B}_0  \nonumber  \\
 &  &  + \,\, {\cal O} ( B_{00} \epsilon^2 / \ell_\perp ) \nonumber \\
 & = & B_0 \left[ \hat{\bf B}_0 \cdot \nabla ( \nabla_\perp h ) +
(2 H_B)^{-1} \nabla_{\perp} h + \alpha \hat{\bf B}_0 \right] \nonumber \\
 & & + \,\, {\cal O} ( B_{00} \epsilon^2 / \ell_\perp ) , \label{eq:curlB}
\end{eqnarray}
where $\alpha$ is defined by
\begin{equation}
\alpha ({\bf r},t) \equiv - \nabla_\perp^2 h .
\label{eq:alpha}
\end{equation}
Inserting expression (\ref{eq:curlB}) into equation (\ref{eq:dvdt}),
we find for the acceleration perpendicular to ${\bf B}_0$:
\begin{displaymath}
\left( \frac{\partial {\bf  v}} {\partial t} + {\bf v} \cdot \nabla
{\bf v} \right)_\perp =
\end{displaymath}
\begin{equation}
 v_A^2 \left\{ \left[ \hat{\bf B}_0 \cdot
\nabla ( \nabla_\perp h ) + (2 H_B)^{-1} \nabla_\perp h \right]
\times \hat{\bf B}_0 + \alpha \nabla_\perp h \right\} ,
\label{eq:dvdt1}
\end{equation}
where $v_A (z) \equiv B_{00} / \sqrt{4 \pi \rho_0}$. We now consider
the parallel component of the vorticity equation. Using equation
(\ref{eq:vperp2}), the parallel vorticity can be approximated as
\begin{equation}
\omega \equiv \hat{\bf B}_0 \cdot ( \nabla \times {\bf v} ) 
\approx - \nabla_\perp^2 f ,
\label{eq:omega}
\end{equation}
and using ${\bf v} \cdot \nabla {\bf v} = \nabla ( \onehalf v^2 ) +
( \nabla \times {\bf v} ) \times {\bf v}$, the inertial term in
equation (\ref{eq:dvdt1}) becomes
\begin{equation}
\hat{\bf B}_0 \cdot \left[ \nabla \times ( {\bf v} \cdot \nabla
{\bf v} ) \right] \approx \hat{\bf B}_0 \cdot ( \nabla_\perp \omega
\times \nabla_\perp f ) .
\end{equation}
The cross-product on the right-hand side of equation (\ref{eq:dvdt1})
yields
\begin{equation}
\hat{\bf B}_0 \cdot \left[ \nabla \times \left\{ \left[ \hat{\bf B}_0
\cdot \nabla ( \nabla_\perp h ) + (2 H_B)^{-1} \nabla_\perp h \right]
\times \hat{\bf B}_0 \right\} \right] = \hat{\bf B}_0 \cdot \nabla
\alpha ,
\end{equation}
where equation (\ref{eq:dB0hat}) is used several times, and $H_B$
eventually drops out of the expression. Therefore, we obtain the
following scalar form of the vorticity equation:
\begin{equation}
\frac{\partial \omega} {\partial t} + \hat{\bf B}_0 \cdot
( \nabla_\perp \omega \times \nabla_\perp f ) = v_A^2 \left[
\hat{\bf B}_0 \cdot \nabla \alpha + \hat{\bf B}_0 \cdot
( \nabla_\perp \alpha \times \nabla_\perp h ) \right] .
\label{eq:dodt2}
\end{equation}
Here $v_A$ depends on position $z$ along the flux tube, and
$\hat{\bf B}_0$ depends on $x$ and $y$ as described by equation
(\ref{eq:dB0hat}). The magnetic field strength $B_0$ is nearly
constant over the cross-section of the tube and can be approximated by
its value on axis, $B_0 \approx B_{00} (z)$. Therefore, in the
remainder of this paper we will write the magnetic field strength as
$B_0 (z)$.

The last step is to replace $\nabla_\perp$ with $\nabla_x \equiv
\nabla - \hat{\bf z} ( \hat{\bf z} \cdot \nabla)$ in the definitions
of $\alpha$ and $\omega$, and in the nonlinear terms of equations
(\ref{eq:dhdt2}) and (\ref{eq:dodt2}), but not in the linear terms
where quantities are differentiated {\it along} the background field.
Then $\nabla_\perp^2 \approx \partial^2 / \partial x^2 + \partial^2 /
\partial y^2$, and the dynamical equations (\ref{eq:dhdt2}) and
(\ref{eq:dodt2}) can be written as
\begin{eqnarray}
\frac{\partial h} {\partial t} & = & \hat{\bf B}_0 \cdot \nabla f
+ \frac{f} {H_B} + [ f , h ] , \label{eq:dhdt} \\
\frac{\partial \omega} {\partial t} & = & - [ \omega , f ] + v_A^2
\left\{ \hat{\bf B}_0 \cdot \nabla \alpha + [ \alpha , h ] \right\} ,
\label{eq:dodt}
\end{eqnarray}
where $[ \cdots , \cdots ]$ is the bracket operator:
\begin{equation}
[a,b] \equiv
\frac{\partial a} {\partial x} \frac{\partial b} {\partial y} -
\frac{\partial a} {\partial y} \frac{\partial b} {\partial x} .
\label{eq:bracket}
\end{equation}
We also derive an alternative form of the dynamical equations.
Taking the second derivative of equation (\ref{eq:dhdt}), we find
\begin{displaymath}
\frac{\partial \alpha} {\partial t}   =   - \nabla_\perp^2 (
\hat{\bf B}_0 \cdot \nabla f + [ f , h] ) + \frac{\omega} {H_B}
\end{displaymath}
\begin{equation}
  =  \hat{\bf B}_0 \cdot \nabla \omega + [ \omega, h] +
[ f , \alpha ] - 2 \left[ \frac{\partial f} {\partial x} ,
\frac{\partial h} {\partial x}  \right] -2 \left[ \frac{\partial f}
{\partial y} , \frac{\partial h} {\partial y}  \right] ,
\label{eq:dadt}
\end{equation}
where we used equation (\ref{eq:dB0hat}) to evaluate the horizontal
derivatives of $\hat{\bf B}_0$ (note that $H_B$ is eliminated from the
equation). In analogy with the Elsasser variables \citep[][]
{Elsasser1950}, we define
\begin{equation}
f_\pm \equiv f \pm v_A h , ~~~~ \mbox{and} ~~~~
\omega_\pm \equiv \omega \pm v_A \alpha , \label{eq:pm}
\end{equation}
and inverting these relations, we can write $f = (f_+ + f_-)/2$,
$h = (f_+ - f_-)/(2 v_A)$, $\omega = (\omega_+ + \omega_-)/2$ and
$\alpha = ( \omega_+ - \omega_- )/(2 v_A )$. Therefore, combining
equations (\ref{eq:dodt}) and (\ref{eq:dadt}), we obtain the following
equations for $\omega_\pm$:
\begin{equation}
\frac{\partial \omega_\pm} {\partial t} = \pm v_A ~ \hat{\bf B}_0
\cdot \nabla \omega_\pm - v_A \frac{d v_A} {dz} \alpha
- [ \omega_\pm , f_\mp ] \pm {\cal N} ,
\label{eq:dodt_pm}
\end{equation}
where the nonlinear term ${\cal N}$ is defined by
\begin{equation}
{\cal N} \equiv \left[ \frac{\partial f_+} {\partial x} ,
\frac{\partial f_-} {\partial x} \right] + \left[
\frac{\partial f_+} {\partial y} , \frac{\partial f_-}
{\partial y}  \right] . \label{eq:N}
\end{equation}
The equations for $\omega_+$ and $\omega_-$ describe inward and
outward propagating Alfv\'{e}n waves, respectively ($\mp \hat{\bf B}_0$
directions); the term with $dv_A/dz$ describes linear coupling between
these waves; and the last two terms describe nonlinear coupling.
In Appendix A we demonstrate that equation (\ref{eq:dodt_pm}) is
consistent with earlier formulations of the wave transport equations
based on the Elsasser variables. The numerical methods used in solving
these equations are discussion in section \ref{sect:Numerics} and
Appendix B.

\subsection{Background Atmosphere}
\label{sect:Background}

We now describe the undisturbed state of the flux tube before any
waves are injected. In the closed field case there are two ``lower
atmospheres'' located at each end of the flux tube. The
chromosphere-corona TRs are located at positions $z = z_{TR}$ and
$z = L_{cor} + z_{TR}$, where $z_{TR}$ is the height of the first TR,
and $L_{cor}$ is the coronal loop length. The total length of the tube
is $L = L_{cor} + 2 z_{TR}$. The gas pressure $p_0 (z)$, density
$\rho_0 (z)$ and magnetic field strength $B_0 (z)$ are functions of
$z$ only, i.e., we neglect variations of these quantities in planes
perpendicular to the flux tube axis. Furthermore, the functions
$p_0 (z)$, $\rho_0 (z)$, $R(z)$ and $B_0 (z)$ are assumed to be
symmetric with respect to the mid-point of the loop ($z = L/2$),
so the two halfs of the loop are identical.

The structure of the lower atmosphere is based on a model of a facular
region (i.e., very bright plage region) called model P, which was
first developed by \citet[][]{Fontenla1999} and more recently
discussed by \citet{Fontenla2006, Fontenla2009}. The temperature
$T_0 (z)$ and molecular
weight $\mu_0 (z)$ are specified as function of height to be in rough
agreement with this model. At the base of the photosphere, we assume
$B_0 (0) = B_{phot} = 1400$ [G] and $p_0 (0) = 3 \times 10^{4}$
$\rm dyne ~ cm^{-2}$, which is the {\it internal} pressure of the flux
tube; then the total pressure ($p_0 + B_0^2/8\pi$) is about $1.1
\times 10^{5}$ $\rm dyne ~ cm^{-2}$, consistent with model P.
The gas pressure $p_0 (z)$ and density $\rho_0 (z)$ as functions of
height are determined by solving a modified hydrostatic equilibrium
equation, $dp_0/dz = - \rho_0 g_{eff}$, where $g_{eff}$ is the
gravitational acceleration corrected for the effects of turbulent
motions. The pressure at the TR is given by
\begin{equation}
p_0 (z_{TR}) = 1.833 \exp \left( - \frac{z_{TR} - 1.8} {0.3278}
\right) ~~~~ {\rm dyne ~ cm^{-2}} , \label{eq:ptr}
\end{equation}
where $z_{TR}$ is in Mm. The magnetic field
strength in the lower atmosphere is given by
\begin{equation}
B_0 (z) = \left[ ( B_{phot}^2 - B_{cor}^2 ) \frac{p_0(z)-p_0(z_{TR})}
{p_0 (0)-p_0(z_{TR})} + B_{cor}^2 \right]^{1/2}  \mbox{for} ~
0 \le z \le z_{TR} ,
\end{equation}
where $B_{phot} = 1400$ G, and $B_{cor}$ is the field strength at the
coronal base (i.e., at $z=z_{TR}$). In the photosphere $p_0 \gg
p_0(z_{TR})$ and $B_0 \gg B_{cor}$, so that the ratio of gas- and
magnetic pressures is approximately constant:
\begin{equation}
\beta (z) \equiv \frac{8 \pi p_0 (z)} {B_0^2 (z)} \approx 0.4 .
\end{equation}
At larger heights in the chromosphere $B_0(z)$ approaches the coronal
value, and $\beta (z)$ decreases with height to values well below
unity.

In the TR the temperature $T_0 (z)$ increases rapidly with height from
about $10^4$ K in the upper chromosphere to about $10^6$ K in the
corona. We do not attempt to resolve this temperature structure, and
treat it as a discontinuity. For closed loops we use the following
profile for the temperature in the corona:
\begin{equation}
T_0 (z) = T_{max} \left[ 1 - 0.8 u^2 (z) \right]^{2/7} ,
\end{equation}
where $u(z) \equiv -1 + 2(z-z_{TR})/L_{cor}$, which lies in the range
$-1 \le u \le +1$, and $T_{max}$ is the peak temperature in the loop as
predicted by the RTV scaling law, equation (\ref{eq:Tmax}). It follows
that $T_0 (z_{TR}) = 0.631~T_{max}$, so the temperature in the upper
TR is a fixed fraction of the peak temperature in the corona. The
above profile is similar to that predicted for a conduction-dominated
loop with constant cross-section and uniform heating \citep[e.g.,][]
{Vesecky1979, Martens2000, Martens2010}. Although it is not difficult
to include the effects of gravity \citep[e.g.,][]{Serio1981}, in this
paper we neglect gravity in the corona, so that the gas pressure
$p_0 (z)$ is constant along the coronal part of the loop. The pressure
is continuous across the TR, so the coronal pressure $p_{cor} = p_0
(z_{TR})$. Therefore, the coronal pressure is determined by the height
$z_{TR}$ of the TR. The plasma density in the corona is given by
\begin{equation}
\rho_0 (z) = \rho_0 (z_{TR}) \left[ \frac{1 - 0.8 u^2 (z)} {0.2}
\right]^{-2/7} , \label{eq:rho0}
\end{equation}
where $\rho_0 (z_{TR})$ is the density at the coronal base, which
is computed from $p_{cor}$ and $T_0(z_{TR})$. For closed loops the
coronal field is approximated as
\begin{equation}
B_0 (z) = B_{cor} \left\{ 1 + (\Gamma -1) [1 - u^2 (z) ] \right\}^{-1} ,
\label{eq:B0}
\end{equation}
where $\Gamma$ is the areal expansion factor; for open fields we use
$B_0 (z) = B_{cor} =$ constant in the corona. In either case the
magnetic flux $\Phi$ ($= \pi R^2 B_0$) is constant along the flux
tube, so the tube radius is given by
\begin{equation}
R(z) = R_{phot} \sqrt{B_{phot}/B_0(z)} .
\end{equation}
Here $R_{phot} = 100$ km is the tube radius at the base of the
photosphere ($z=0$ and $z=L$). The radius $R$ in the corona depends on
the parameters $B_{cor}$ and $\Gamma$, and is different for the
different models.

\begin{figure*}
\epsscale{1.09}
\plotone{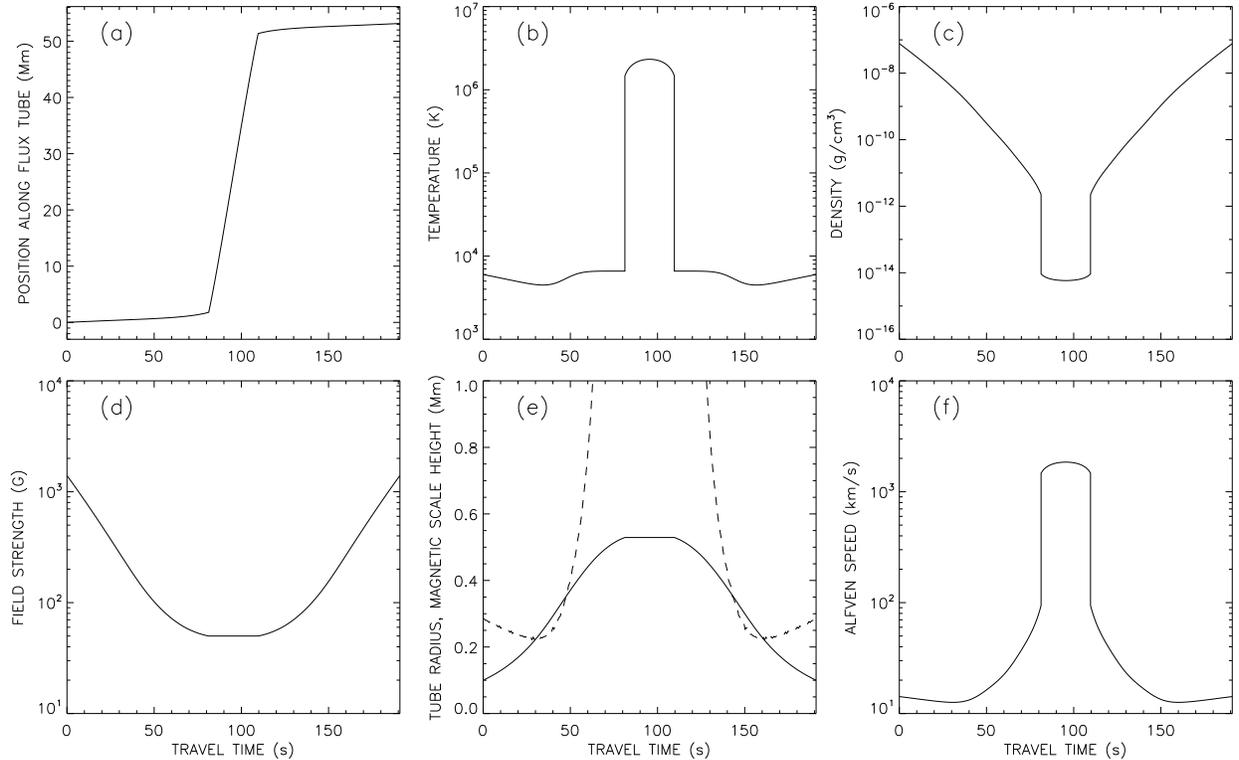}
\caption{%
Reference model for a coronal loop and the two ``lower atmospheres''
at the two ends of the loop. Various quantities are plotted as
function of the Alfv\'{e}n wave travel $\tau$ to a given point along
the loop: (a) position $z (\tau)$ along the loop as measured from the
left footpoint; (b) temperature $T_0$; (c) mass density $\rho_0$;
(d) magnetic field strength $B_0$; (e) flux tube radius $R$ ({\it
full} curve) and magnetic scale height $| H_B |$ ({\it dashed} curve);
(f) Alfv\'{e}n speed $v_A$. The two chromosphere-corona TRs are
located at $\tau = 81.3$ [s] and $\tau = 109.6$ [s].}
\label{fig:figure1}
\end{figure*}
We consider a reference model with coronal field strength $B_{cor} =
50$ G, expansion factor $\Gamma = 1$, and TR height $z_{TR} = 1.8$ Mm,
which corresponds to a coronal pressure $p_{cor} = 1.833$
$\rm dyne ~ cm^{-2}$. The coronal loop length $L_{cor} = 49.6$ Mm, which
is the closest one can get to 50 Mm with the chosen time step $\Delta
t_0 = 0.746$ s (see Appendix B for details). The total
loop length is $L = 53.6$ Mm. Figure \ref{fig:figure1} shows various
quantities plotted as function of position along the flux tube for
this model. Positions are given in terms of the Alfv\'{e}n wave travel
time from the left footpoint ($z=0$):
\begin{equation}
\tau (z) \equiv \int_0^z \frac{dz} {v_A (z)} . \label{eq:tau}
\end{equation}
Figure \ref{fig:figure1}a shows the relationship between $z$ and
$\tau$ for the reference model; the photospheric footpoints are
located at $\tau(0)=0$ and $\tau(L)= 190.9$ s, and the corona is
located in the region $81.3 < \tau < 109.6$ s. The other panels in
Figure \ref{fig:figure1} show the temperature $T_0$, density
$\rho_0$, magnetic field strength $B_0$, flux tube radius $R$, the
absolute value of the magnetic scale height $H_B$ (defined in equation
(\ref{eq:HB})), and the Alfv\'{e}n speed $v_A$ as functions of
Alfv\'{e}n travel time $\tau$ from the left footpoint. Note that when
expressed in terms of $\tau$, the corona is only a small part of the
computational domain.

Figure \ref{fig:figure1}e shows that $|H_B| < R$ in two intervals,
$30 < \tau < 48$ [s] and $140 < \tau < 158$ [s], which correspond to
the temperature minimum regions at the two ends of the loop. In these
regions the quantity $\epsilon_0$ defined in equation (\ref{eq:eps0})
is greater than unity, so the thin tube approximation is no longer
valid. Clearly, the thin tube and reduced MHD approximations discussed
in section \ref{sect:RMHD} provide only a crude description of the
magnetic structure and wave dynamics in the lower atmosphere. A proper
treatment will require full MHD simulations, and is beyond the scope
of the present work.

The present model predicts braiding of the coronal field lines on a
transverse scale less than the tube radius, $R_{cor} \approx 529$ km.
This radius is less than the resolution limits of present-day X-ray
telescopes. Therefore, we should expect that the predicted coronal
structures will be difficult to observe.

\subsection{Photospheric Footpoint Motions}
\label{sect:Footpoints}

The Alfv\'{e}n waves are produced by footpoint motions imposed at the
two ends of the flux tube, $z = 0$ and $z = L$.
In reality these motions may distort the shape of the flux tube as
indicated in Figure \ref{fig:cartoon}, but here we use a simpler
approach in which the motions are assumed to be confined to a circular
area $x^2+y^2 \le R_{phot}^2$ at $z=0$, and similar at $z=L$. The
velocity ${\bf v} (x,y,0,t)$ at these boundaries can be written in
terms of polar coordinates $(r,\varphi)$, where $r$ is the distance
from the flux tube axis and $\varphi$ is the azimuth angle in the
$(x,y)$-plane. We assume that the radial component of velocity
vanishes at the tube wall, $v_r (R_{phot},\varphi,0,t) = 0$, so that
the circular shape of the cross-section is preserved. We also assume
that the motions are horizontal and incompressible:
\begin{equation}
{\bf v} (r,\varphi,0,t) = \nabla f \times \hat{\bf z} ,
\end{equation}
where $f(r,\varphi,0,t)$ is the stream function at $z=0$, and similar
at $z=L$.
As described in Appendix B, functions on a circular area can be
decomposed into orthogonal basis functions $F_i (\xi, \varphi)$,
where $\xi$ is the relative distance from the tube axis ($\xi \equiv
r/R$) and index $i$ enumerates the basis functions ($i = 1, \cdots ,
N$). We use a relatively small number of basis functions ($N=92$);
the functions are shown in Figure \ref{fig:func2}.

In this paper we assume that the footpoint motions have a pattern
consisting of two counter-rotating cells. This pattern can
be described as a superposition of two modes with azimuthal mode
number $m=1$. For the particular set of basis functions used in the
present paper, these driver modes have indices $i = 7$ and $i = 8$
(see Appendix B for details), and are shown in the top row of
Figure \ref{fig:func2} (seventh and eighth image from the left).
Both modes have the same {\it radial} dependence given by
$J_0 ( a_\perp r/R_{phot})$, where $J_0 (x)$ is the zeroth order
Bessel function of the first kind, and $a_\perp$ is the dimensionless
perpendicular wavenumber, which equals 3.832 for these particular
modes (the first zero of the Bessel function).  However, the {\it
azimuth} dependence of the two modes is different: the mode with $i=7$
is proportional to $\cos \varphi$, while the one with $i = 8$ is
proportional to $\sin \varphi$. The imposed stream function at $z=0$
can then be written as a superposition of the two modes:
\begin{equation}
f(r,\varphi,0,t) = f_7(t) F_7 (r/R_{phot},\varphi) +
f_8(t) F_8 (r/R_{phot},\varphi) .
\end{equation}
Here $f_7(t)$ and $f_8(t)$ are the mode amplitudes, which vary
randomly with time $t$ and are not correlated with each other.
The vertical component of vorticity at $z=0$ is then given by
\begin{equation}
\omega (r,\varphi,0,t) = - \nabla_\perp^2 f = (3.832/R_{phot})^2
f(r,\varphi,0,t) .
\end{equation}
The above time-dependent pattern simulates the intermixing of the
plasma within the flux tube due to motions imposed by the surrounding
convective flows. 

The random variables $f_7 (t)$ and $f_8 (t)$ are constructed as
follows. For each variable, we first create a normally distributed
random sequence $f(t)$ on a grid of times covering the entire length
of the simulation ($t_{max} = 3000$ s). Then the sequence is filtered
in the Fourier domain using a Gaussian function, $G(\tilde\nu) =
\exp [-(\tau_0 \tilde\nu)^2]$, where $\tilde\nu$ is the temporal
frequency (in Hz) and $\tau_0$ is the specified correlation time
(for the reference model $\tau_0 = 60$ s). The filtered sequence is
renormalized such that the rms vorticity of each mode equals a
specified value, $\omega_0$. The root-mean-square (rms) velocity of
the combined pattern of the two modes is $\Delta v_{rms} = \sqrt{2}
R_{phot} \omega_0 / 3.832$. The quantities $\tau_0$ and $\omega_0$ are
free parameters of the model.

\subsection{Numerical Methods}
\label{sect:Numerics}

The techniques used for solving the RMHD equations are described
in Appendix B, and the energy equation is discussed in Appendix C.
The transverse dependence of the waves is
described using a spectral method, i.e., all scalar functions are
written as sums over 92 discrete modes. The nonlinear terms in the
equations are represented by a matrix $M_{kji}$ that couples the
different modes, and cause transfer of energy from low to high
wavenumber. The modes with the highest transverse wavenumbers are
artificially damped, which describes viscous and resistive processes
on small spatial scales. The damping rate is given by equation
(\ref{eq:nuk}) with $\nu_{max} = 0.7$ $\rm s^{-1}$.
For the $z$-dependence of the waves we use finite differences;
for example, in the model shown in Figure \ref{fig:figure1} there are
259 points along the tube. To accurately simulate the wave
propagation, we use a grid that is uniform in Alfv\'{e}n wave travel
time $\tau (z)$ with a grid spacing $\Delta \tau$ equal to the time
step $\Delta t_0$ of the simulation (for the reference model, $\Delta
t_0 = 0.746$ s). At the chromosphere-corona TR the density $\rho_0
(z)$ is discontinuous, which is represented by two grid points at the
same position $z_{TR}$ but with different densities. The discontinuity
causes wave reflections that are described in terms of reflection and
transmission coefficients (see Appendix B for details).

The RMHD model is valid only when $\Delta B_{rms} /B_0 \ll 1$, where
$\Delta B_{rms}$ is the rms value of the transverse magnetic field
fluctuation. We will find that this condition is only marginally
satisfied. Therefore, the present model can describe only some of the
dynamical processes that occur in the chromosphere and corona.
In particular, the model does not describe field-aligned flows such as
spicules.

\subsection{Limitations of the Model}
\label{sect:Limitations}

The present model for chromospheric and coronal heating has several
drawbacks and limitations. The code uses an explicit numerical
scheme, which makes it difficult to simulate waves in models with
high coronal field strength. On the real Sun the field strength in
active region loops in the low corona is 100 - 500 G, but in the
present paper we must limit ourselves to $B_{cor} \le 200$ G. Also,
the model includes only a limited number of wave modes (see Figure
\ref{fig:func2}), and the time step is relatively large ($\Delta
t_0 > 0.1$ s), so the turbulent spectrum is not well resolved.

The model considers only a single magnetic flux tube, and does not
allow for splitting and merging of magnetic flux elements at the two
ends of the coronal loop \citep[][] {Berger1996}. It is unclear to
what extent the chromospheric and coronal heating rates and the
spatial distribution of the heating are affected by this
approximation. Perhaps more importantly, the present RMHD model does
not allow for feedback of the heating on the background atmosphere.
In the present version there are no field-aligned flows, and the
temperature and density are assumed to be constant in time, so we do
not simulate the dynamic response of the atmosphere to heating
events. This also means that we cannot make predictions regarding the
temporal variability of EUV and X-ray emissions from the modeled
coronal loops. In particular, we do not simulate spicules or similar
dynamic events, even though such events may in fact be driven by
nonlinear Alfv\'{e}n waves \citep[see][]{Matsumoto2010}. The plasma
density in our model is assumed to be constant in planes perpendicular
to the tube axis, so there are no variations in density on scales less
than the tube radius. Therefore, the present model has little to say
about the multi-thermal structure of coronal loops.

Finally, the model does not include the effects of phase mixing
\citep[][]{Heyvaerts1983} or resonant absorption \citep[e.g.,][]
{DeGroof2002}, which depend on the variation of Alfv\'{e}n speed
in the plane perpendicular to the tube axis. In the present model
both $B_0$ and $\rho_0$ are constant in the $(x,y)$ plane.

\section{Results}

\subsection{Reference Model}
\label{sect:RefMod}

We first construct a reference model with coronal field strength
$B_{cor} = 50$ G, expansion factor $\Gamma = 1$, and coronal loop
length of about $50$ Mm. More precisely, the coronal loop length
$L_{cor} = 49.6$ Mm, the transition-region height $z_{TR} = 1.8$ Mm,
and the total tube length $L = 53.2$ Mm. The footpoint motions
have a correlation time $\tau_0 = 60$ s, each of the driver modes has
a vorticity $\omega_0 = 0.04$ $\rm s^{-1}$, and the rms velocity is
$\Delta v_{rms} = 1.48$ $\rm km ~ s^{-1}$. The latter is reasonable
compared to the convective velocities of several km/s expected to
exist in the downflow lanes below the photosphere. The TR height
corresponds to a coronal pressure $p_{cor} = 1.833$ $\rm dyne ~
cm^{-2}$, which is typical for hot loops found in active regions, and
equation (\ref{eq:Tmax}) yields a peak temperature $T_{max} = 2.3$ MK.

\begin{figure*}
\epsscale{0.96}
\plotone{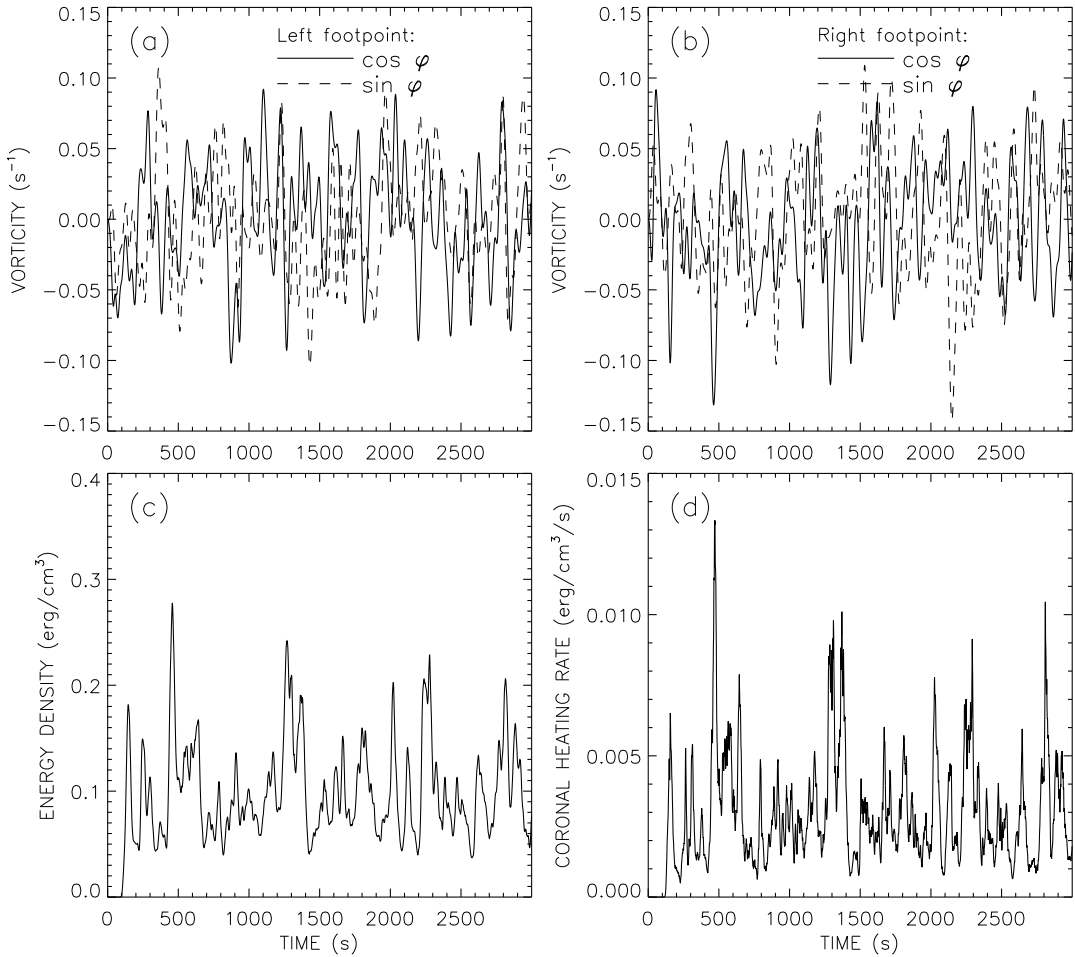}
\caption{Various quantities plotted as function of time for the
reference model:
(a) imposed vorticity $\omega_k$ at the left footpoint ($z=0$)
for $k=6$ ({\it full} curve) and $k=7$ ({\it dashed} curve);
(b) similar for the right footpoint ($z=L$);
(c) energy density $E$ in the corona;
(d) coronal heating rate, $Q_{cor}$.}
\label{fig:figure2}
\end{figure*}
The vorticities of the driver modes are shown as function of time
in Figures \ref{fig:figure2}a and \ref{fig:figure2}b for the left
($z=0$) and right ($z=L$) footpoints, respectively. These random
footpoint motions create Alfv\'{e}n waves that propagate upward along
the flux tube. Initially, only the two driver modes are present. The
decrease of density with height causes the velocity amplitude of the
waves to increase with height. In the chromosphere, part of the wave
energy is reflected back down due to the increase of Alfv\'{e}n speed
with height. Even stronger reflection occurs at the
chromosphere-corona TR, where the Alfv\'{e}n speed suddenly increases
by about a factor 15. These reflections produce a pattern of
counter-propagating waves in the photosphere and chromosphere at the
two ends of the loop. The wave amplitudes in the chromosphere soon
build up to about 20 - 30 km/s, consistent with the observations by
\citet{DePontieu2007b}.

Due to evolution of the driver modes and time delays in the
reflection, the inward and outward propagating waves at a given height
$z$ have different spatial distributions in the $(x,y)$ plane.
Furthermore, these patterns are superpositions of multiple modes with
different perpendicular wavenumbers. Such counter-propagating waves
with different spatial patterns and multiple modes are subject to
strong nonlinear interactions \citep[e.g.,][]{Shebalin1983,
Higdon1984, Oughton1995, Goldreich1995, Goldreich1997,
Bhattacharjee2001, Cho2002, Oughton2004}. In our RMHD model,
these interactions are due to the bracket terms in equations
(\ref{eq:dhdt}), (\ref{eq:dodt}) and (\ref{eq:dodt_pm}).
Basically, the inward
propagating waves are distorted by the outward propagating waves,
and {\it vice versa} \citep[e.g.][]{Chandran2009}. These distortions
are large because the fluid displacements are comparable to the
transverse scale of the waves. For example, for chromospheric waves
with velocity amplitude of 10 km/s acting over a period of 50 s, the
transverse displacement is about 500 km, equal to the transverse scale
of the driver modes.  As a result of these nonlinear interactions,
other wave modes with smaller spatial scales are excited (see Figure
\ref{fig:func2} for a display of the 92 wave modes used in the present
model). After about 200 s from the start of the simulation, a
well-developed spectrum of Alfv\'{e}n waves has formed and dissipation
of the high-wavenumber waves becomes significant. This dissipation is
due to a combination of viscous and resistive diffusion effects,
so the model includes the effects of magnetic reconnection. Such
turbulent dissipation of waves in the lower atmosphere continues
throughout the simulation.

A small fraction of the wave energy is transmitted through the TR
into the corona. Energy is injected into the coronal loop at both
ends, producing counter-propagating waves in the corona. As in the
chromosphere, the counter-propagating waves significantly distort each
other because (1) they have different spatial patterns and (2) the
fluid displacements are comparable to the transverse scale of the
waves. Therefore, the waves produce turbulence in the corona, and
there is a continual cascade of energy to smaller transverse scales.
As a result, part of the Alfv\'{e}n wave energy is dissipated in the
coronal part of the loop, again due to combination of viscous and
resistive effects. We find that the rms velocity amplitude of the
waves in the corona is similar to that in the upper chromosphere,
but the energy dissipation rate in the corona is much lower because of
the lower coronal density.

In the present model, all of the dissipation occurs via Alfv\'{e}n
wave turbulence, both in the chromosphere and in the corona. Phase
mixing and resonant absorption of Alfv\'{e}n waves \citep[e.g.,][]
{Heyvaerts1983, Poedts1989, Ofman1994, Erdelyi1995, DeGroof2002} play
no role because these processes require variations in Alfv\'{e}n speed
across the field lines, which are not included in the present model.
A key feature of the model is that the photospheric footpoint motions
include more than one driver mode, not just the torsional mode as in
the simulations by \citet{Antolin2010} and \citet{Matsumoto2010}.
Specifically, we use two driver modes (modes $k=7$ and $k=8$ in
Figure \ref{fig:func2}) with azimuthal mode number $m=1$. If the
system is driven using only a single mode (e.g., $k=7$), all
nonlinear terms in equations (\ref{eq:dodt}) and (\ref{eq:dhdt})
vanish.  In this case all energy remains in the driver mode, and
no turbulence develops, as we have verified in test calculations.
Hence, for Alfv\'{e}n wave turbulence to develop it is essential
that the footpoint motions include multiple driver modes that have
nonlinear couplings with high wavenumber modes. This coupling can
be indirect; for the chosen $m=1$ modes, the coupling runs via the
$m=0$ modes.

Figure \ref{fig:figure2}c shows the spatially averaged energy density
of the waves $E(t)$ in the corona as function of time (average between
$z = z_{TR}$ and $z = z_{TR} + L_{cor}$). This quantity is the sum of
magnetic free energy $E_{mag}(t)$ and kinetic energy $E_{kin}(t)$, but
the magnetic energy dominates. Figure \ref{fig:figure2}d shows
the spatially averaged heating rate $Q_{cor}(t)$ per unit volume in
the corona. Note that both the energy density and the heating rate
vary strongly with time. Significant heating starts only after about
200 seconds when the waves have become sufficiently turbulent.
During the latter part of the simulation, the average heating rate in
the coronal part of the loop is $2.98 \times 10^{-3}$ $\rm erg ~
cm^{-3} ~ s^{-1}$. This is consistent with the second RTV scaling law,
equation (\ref{eq:QcorRTV}), which yields $Q_{cor} = 2.9 \times
10^{-3}$ $\rm erg ~ cm^{-3} ~ s^{-1}$. Therefore, the reference model
describes a coronal loop in which the average rate of plasma heating
is balanced by radiative and conductive losses. The model represents
the kind of hot, dense loops typically found in active regions. We
conclude that such loops can be heated by Alfv\'{e}n wave turbulence
generated inside the magnetic flux tubes by footpoint motions with rms
velocity of about 1.5 km/s.

\begin{figure*}
\epsscale{1.09}
\plotone{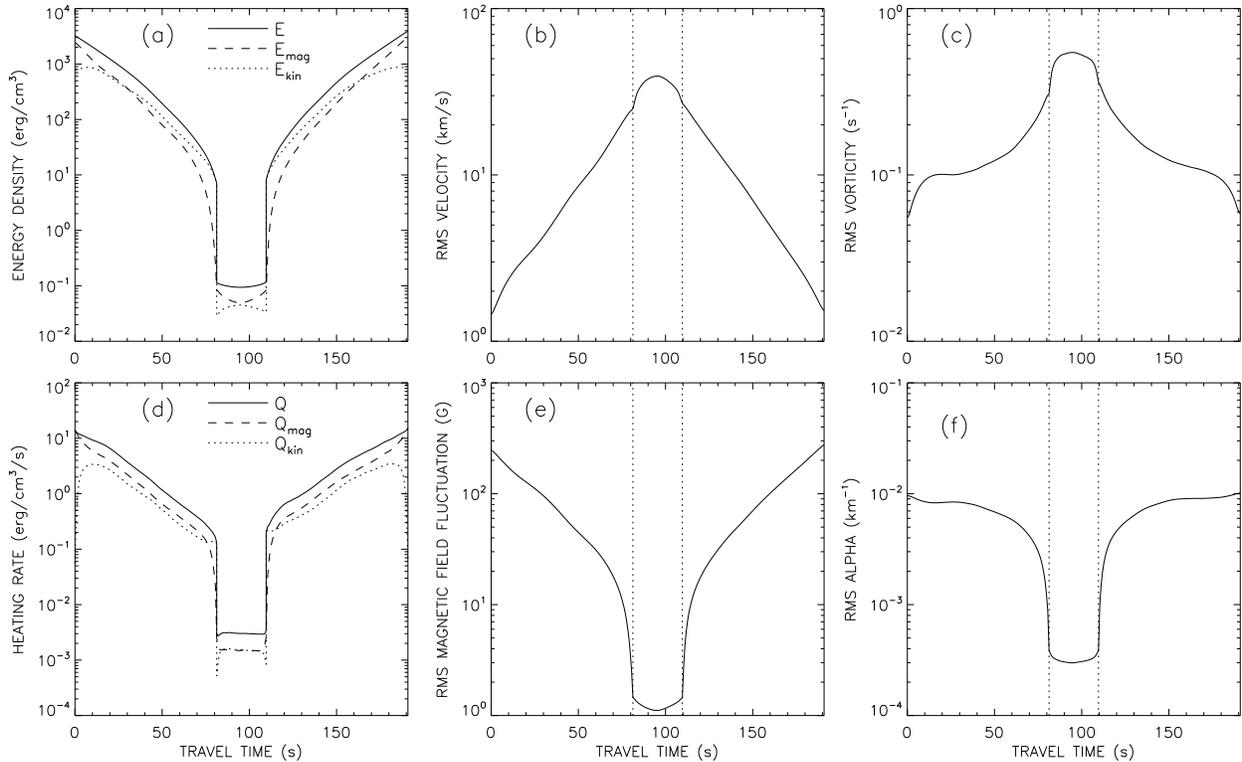}
\caption{Model for Alfv\'{e}n wave turbulence in a coronal
loop. Various quantities are plotted as function of position along the
flux tube for the reference model: (a) kinetic and magnetic energy
densities, and their sum $E(z)$; (b) rms velocity $\Delta v_{rms}$;
(c) rms vorticity $\omega_{rms}$, (d) kinetic and magnetic heating
rates, and their sum $Q(z)$; (e) rms magnetic fluctuation $\Delta
B_{rms}$; (f) rms twist parameter $\alpha_{rms}$. These quantities are
averaged over the cross-section of the flux tube and over time.
Positions along the loop are given in terms of the Alfv\'{e}n travel
time $\tau (z)$. The corona is located in the range $81.3 < \tau <
109.6$ s, which is indicated by vertical {\it dotted} lines.}
\label{fig:figure3}
\end{figure*}

Figure \ref{fig:figure3} shows various quantities as function of
position along the flux tube, averaged over the cross-section of the
flux tube ($x$ and $y$) and over the time interval $t=[800,3000]$ s.
Positions are given in terms of the Alfv\'{e}n
travel time $\tau (z)$ in seconds. Figure \ref{fig:figure3}a shows the
kinetic and magnetic energy densities, $E_{kin}(z)$ and $E_{mag}(z)$,
and their sum $E(z)$. Figure \ref{fig:figure3}d shows similar plots
for the kinetic and magnetic heating rates, $Q_{kin}(z)$ and
$Q_{mag}(z)$, and their sum $Q(z)$.  These quantities are
discontinuous at the TR. In the deep photosphere ($\tau < 20$ s and
$\tau > 170$ s) and in the corona ($81.3 < \tau < 109.6$ s) the
magnetic energy dominates, but in the chromosphere $E_{kin} >
E_{mag}$; this is similar to what happens in open-field models with
non-WKB wave reflection dominated by low-frequency waves \citep[see
Figure 6 in][]{Cranmer2005}. Despite these differences in wave energy
density, magnetic heating dominates over viscous heating almost
everywhere in the model ($Q_{mag} > Q_{kin}$).

The middle and right panels in Figure \ref{fig:figure3} show the rms
values of the velocity, vorticity, magnetic field fluctuation, and
$\alpha$ parameter. All four of these quantities are continuous
at the TR. The rms velocity $\Delta v_{rms}$ and rms magnetic field
fluctuation $\Delta B_{rms}$ are defined in equations (\ref{eq:DVrms})
and (\ref{eq:DBrms}), and are further averaged over time. Note that
the velocity and vorticity have their peak values at the mid-point of the
loop in the corona, $\Delta v_{rms} \approx 37$ $\rm km ~ s^{-1}$ and
$\omega_{rms} \approx 0.52$ $\rm s^{-1}$ at $\tau = 95.4$ s (see
Figures \ref{fig:figure3}b and \ref{fig:figure3}c). Also note that
the velocity at the mid-point is about 60\% larger than that at the
two TRs (vertical dashed lines in \ref{fig:figure3}b). This is due
to the fact that the waves in the corona are reflected back and forth
between the TRs several times before they decay, creating standing
waves with nodes near the TRs. The velocities of 20 - 40 km/s found
here for the corona are similar to the nonthermal velocities of 
20 - 60 km/s found in observations of spectral line widths in active
regions \citep[e.g.,][]{Dere1993, Warren2008, Li2009} and on the quiet
Sun \citep[][]{Chae1998, McIntosh2008}. This suggests
that the modeled waves have more or less the correct amplitude.
The magnetic fluctuation $\Delta B_{rms} (z)$ and twist parameter
$\alpha_{rms} (z)$ shown in Figures \ref{fig:figure3}e and
\ref{fig:figure3}f have their peaks in the lower atmosphere. This is
due to the large gradients in Alfv\'{e}n speed in the chromosphere and
TR, which cause strong wave reflection and produce a patterns of
nearly standing waves in the lower atmosphere. The non-constancy of
$\alpha_{rms} (z)$ implies that the system is far from a force-free
equilibrium state. The peak value of $\Delta B_{rms} /B_0$ occurs in
the low chromosphere and is about 0.3, so the RMHD approximation is
only marginally satisfied.

\begin{figure*}
\epsscale{1.1}
\plotone{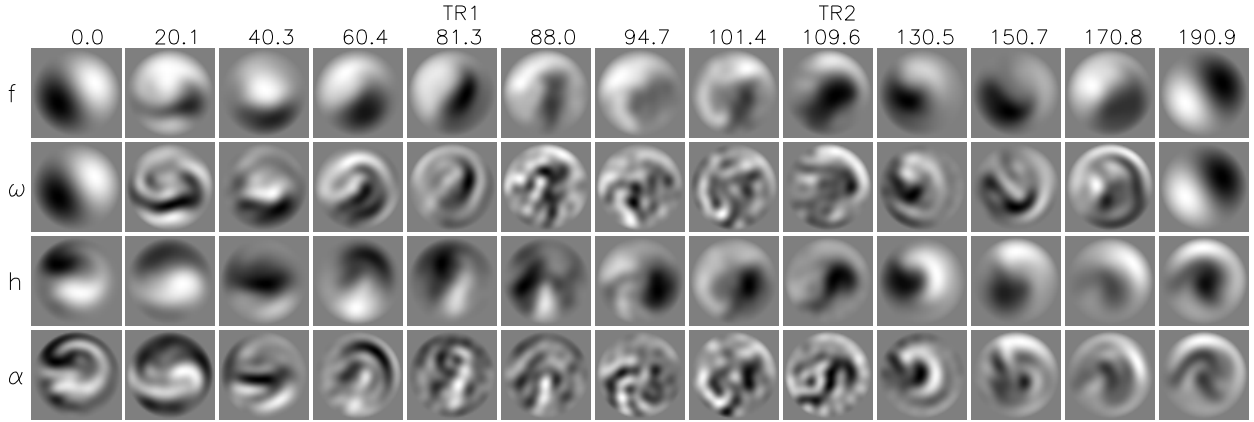}
\caption{Spatial distribution of various dynamical quantities in the
reference model at time $t = 3000$ [s]: velocity stream function
$f$, vorticity $\omega$, magnetic flux function $h$, and twist
parameter $\alpha$.  The different columns correspond to different
positions along the flux tube, and are labeled with the Alfv\'{e}n
travel time $\tau (z)$. Each panel shows the normalized distribution
of the relevant quantity as function of transverse coordinates $x$ and
$y$ (see Figure \ref{fig:figure3} for information about
normalization). A movie sequence is available in the on-line version
of the manuscript.}
\label{fig:figure4}
\end{figure*}
Figure \ref{fig:figure4} shows cross-sections of the tube at various
positions along the loop at the end of the simulation ($t = 3000$ s).
The top row shows the velocity stream function $f(x,y,z)$, the
second row shows the vorticity $\omega (x,y,z)$, the third row shows
the magnetic flux function $h(x,y,z)$ and the bottom row shows the
twist parameter $\alpha (x,y,z)$. The different columns corresponds to
different positions $z$ along the loop, and are labeled with the
Alfv\'{e}n travel time $\tau (z)$ from the left footpoint. The two TRs
are located at $\tau = 81.3$ s and $\tau = 109.6$ s. The upper left
and upper right panels show the footpoint motions that drive the
system. The vorticity $\omega (x,y,z)$ and twist $\alpha (x,y,z)$
exhibit small-scale structures that are produced by nonlinear
interactions, as described above. A movie sequence of such images is
available in the on-line version of the manuscript. This sequence
covers the last 298 seconds of the simulation and shows that the
system is highly turbulent. The waves in the corona have smaller
spatial scales and evolve more rapidly than those in the lower
atmosphere.

\begin{figure*}
\epsscale{0.9}
\plotone{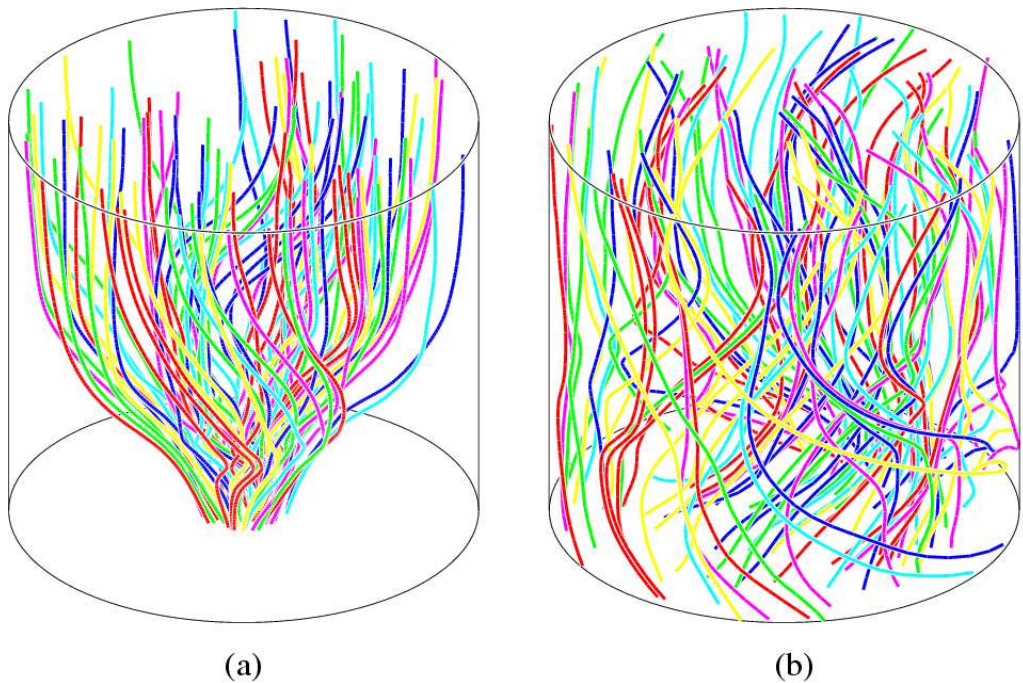}
\caption{Magnetic field lines in the reference model at time $t =
2701.7$ [s], viewed from an angle of $30^\circ$. (a) The lower
atmosphere up to the height of the first TR, $z_{TR} = 2$ Mm. The
starting points of the field lines are randomly distributed inside
the flux tube at height $z=0$ (cylinder base). The radius of cylinder
is $0.53$ Mm, and the vertical scale of the image is compressed by
a factor $1.9$. (b) Continuation of the same field lines
into the coronal part of the loop. The actual length of cylinder is
$49.6$ Mm, so the vertical scale of the image is compressed by a
factor $47.3$. Two movie sequences are available in the on-line
version of the manuscript.}
\label{fig:figure5}
\end{figure*}
Figures \ref{fig:figure5}a and \ref{fig:figure5}b show magnetic field
lines in the reference model at time $t = 2702$ [s] in the lower
atmosphere and in the corona, respectively (the vertical scales of
these images are compressed by different factors). The field lines
are traced from randomly selected points at height $z=0$ in the
photosphere. The field lines are significantly distorted due to the
Alfv\'{e}n waves that travel up and down the flux tube with a range of
transverse wavenumbers. Two movie sequences of such images are
available in the on-line version of the manuscript. These movies show
the evolution of the magnetic field over a period of 298 seconds (from
$t = 2702$ [s] to $t = 3000$ [s]), and are traced from footpoints that
move with the flow. The coronal field lines are to some degree twisted
and braided around each other (see Figure~\ref{fig:figure5}b), but
these structures are highly dynamic and change on a time scale of
seconds. Therefore, the system is not in a force-free state, and is
best described as Alfv\'{e}n wave turbulence. The effects of such
turbulence on the solar corona have been modeled previously for both
open \citep[][]{Hollweg1986, Zhou1990, Matthaeus1999, Dmitruk2001,
Dmitruk2003, Cranmer2003, Cranmer2005, Cranmer2007, Verdini2007} and
closed magnetic fields \citep[][] {Heyvaerts1984, Heyvaerts1992,
Longcope1994, Dmitruk1997, Buchlin2007, Rappazzo2008}. The present
work demonstrates that Alfv\'{e}n wave turbulence can occur both in
the chromosphere and in the corona, and can develop even when the
photospheric footpoint motions occur on very small spatial scales
($\ell_\perp < 100$ km). The model can quantitatively explain the
observed heating rates in active regions.

\subsection{Power Spectra}
\label{sect:Spectra}

The {\it spatial} power spectra for velocity and magnetic field
fluctuations are defined in equation (\ref{eq:Power}). We computed
such spectra for both the standard reference model (described above)
and for a modified version in which the spatial resolution of the
model was slightly increased. Specifically, the maximum perpendicular
wavenumber $a_{max}$ was increased from 20 to 26, which causes the
number of modes to increase from 92 to 158. Also, the maximum
damping rate was increased by a factor $(26/20)^6$, so that the
damping rate $\nu_k$ for the low wavenumber modes ($a_k < 20$) is the
same as that in the standard model [see equation (\ref{eq:nuk})]. We
found that the increased spatial resolution has little effect on the
power in the low wavenumber modes, therefore, the standard model is
adequate for most purposes. Nevertheless, in the following we show
results from the modified version with $a_{max} = 26$.

\begin{figure*}
\epsscale{0.93}
\plotone{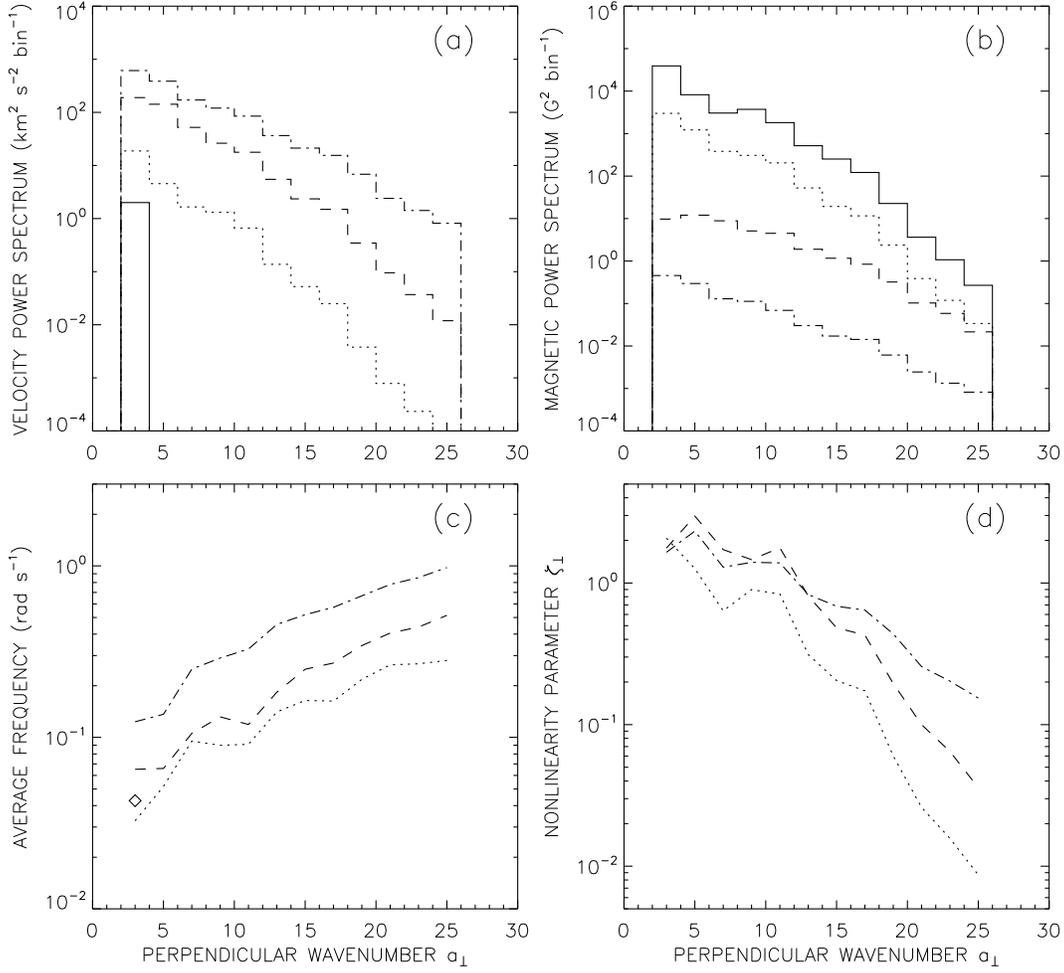}
\caption{Power spectra and related quantities for the reference model:
(a) velocity power spectra, (b) magnetic power spectra, (c) average
frequency $\tilde{\omega}$ of the velocity fluctuations, (d) parameter
describing the degree of nonlinearity of the waves ($\zeta_\perp
\equiv k_\perp v_\perp / \tilde{\omega}$). These quantities are
plotted as function of the dimensionless perpendicular wavenumber
$a_\perp$. The different curves indicate different positions along the
loop: photospheric base ($z = 0$, {\it solid} curves, {\it diamond}),
temperature minimum region ($z = 0.50$ Mm, {\it dotted} curves),
chromosphere ($z = 1.52$ Mm, {\it dashed} curves), corona ($z = 25.2$
Mm, {\it dash-dotted} curves).}
\label{fig:figure6}
\end{figure*}
Figures \ref{fig:figure6}a and \ref{fig:figure6}b show velocity and
magnetic power spectra binned in intervals of the dimensionless
wavenumber $a_\perp$ for four heights in the reference model. 
Specifically, the velocity power in bin $n$ is given by
$\tilde{P}_{V,n} = \sum_k P_{V,k}$, where $P_{V,k}$ is defined in
equation (\ref{eq:Power}) and the sum is taken over all modes with
$a_k$ in the range $n \Delta a < a_k < (n+1) \Delta a$ ($n = 0,
\cdots , 12$). Here $\Delta a = 2$ is the bin size in wavenumber
space. A similar expression holds for the magnetic power spectrum
$\tilde{P}_{B,n}$. The results shown in Figure \ref{fig:figure6} were
derived from the last 800 time steps of the simulation (597 seconds).
The {\it solid} curve in Figure \ref{fig:figure6}a shows the velocity
power spectrum at the base of the photosphere ($z=0$), and is
dominated by the two driver modes with $a_\perp = 3.832$. As we move
to larger heights, the spectrum is broadened and the total power is
increased. At height $z = 1.52$ Mm in the chromosphere, the turbulence
has generated a broad distribution of modes extending up to the
maximum available wavenumber ($a_{max} = 26$). The {\it dash-dotted}
curve in Figure \ref{fig:figure6}a shows the power spectrum near the
mid-point of the coronal loop ($z = 25.2$ Mm) where the level of
turbulence is further enhanced.  Figure \ref{fig:figure6}b shows the
corresponding curves for the magnetic power spectrum. At the base ($z
= 0$, {\it solid} curve) the magnetic spectrum extends over a broad
range of perpendicular wavenumbers and is very different from the
velocity spectrum at that height. The high wavenumber part of this
spectrum is due to the downward propagation of waves produced in the
chromosphere. The magnetic fluctuations in the corona are much smaller
than those in the lower atmosphere at all wavenumbers.

For each wave mode $k$ and height $z$, we can also determine the
{\it temporal} power spectrum $P_k (\tilde{\omega},z)$ of the velocity
fluctuations. Here $\tilde{\omega}$ is the wave frequency in radians
per second. We compute this power spectrum by taking the Fourier
Transform of the velocity stream function $f_k (z,t)$ with respect to
time, and then multiplying the result by the square of the
perpendicular wavenumber, $k_\perp = a_k / R(z)$. The average
frequency $\tilde{\omega}_k$ of the waves can be defined as an
average over the power spectrum:
\begin{equation}
\tilde{\omega}_k (z) \equiv \frac{ \int_0^\infty \tilde{\omega}
P_k (\tilde{\omega},z) d \tilde{\omega} }
{ \int_0^\infty P_k (\tilde{\omega},z) d \tilde{\omega} } .
\end{equation}
We further average these frequencies over modes $k$ to obtain the
average frequency $\tilde{\omega}_n (z)$ for each bin $n$ in
wavenumber space. The three curves in Figure \ref{fig:figure6}c show
$\tilde{\omega}_n$ for three different heights in the reference model.
Note that these average frequencies generally increase with
perpendicular wavenumber, as expected for turbulent flows. The
{\it diamond} symbol in Figure \ref{fig:figure6}c indicates the
average frequency of the footpoint motions.

In fully developed Alfv\'{e}nic turbulence, the fluctuation are
expected to reach a ``critical balance'' in which the average wave
period is of the order of the nonlinear transfer time,
$( k_\perp v_\perp )^{-1}$, where $v_\perp$ is the average velocity.
Following \citet{Goldreich1995}, we define a nonlinearity parameter
\begin{equation}
\zeta_n \equiv \frac{k_{\perp,n} v_{\perp,n}} {\tilde{\omega}_n} ,
\end{equation}
where $v_{\perp,n}$ is the average velocity based on the power in bin
$n$, which is given by $v_{\perp,n}^2 = P_{V,n} a_{\perp,n} / \Delta
a$. Figure \ref{fig:figure6}d shows $\zeta_n$ as function of
dimensionless perpendicular wavenumber. Note that for low wavenumbers
$\zeta_n \sim 1$, indicating the presence of {\it strong} turbulence
at all heights. At large wavenumbers $\zeta_n$ drops below unity,
which is due to wave damping.

These figures demonstrate that the waves injected into the
corona through the TR have a broad range of perpendicular wavenumbers,
and have temporal frequencies that are very different from the driver
modes launched at the base of the photosphere. Therefore, the dynamics
of waves and turbulence in the lower atmosphere is very important for
understanding the coronal heating problem.

\subsection{Open Field Model}
\label{sect:Open}

\begin{deluxetable*}{crcccccc}
\tablewidth{0pt}
\tablecaption{\label{table1}Dependence of Heating Rates on
Footpoint Motions and Coronal Loop Length}
\tablehead{
\colhead{Model} & \colhead{$\tau_0$} & \colhead{$\omega_0$} &
\colhead{$\Delta v_{rms}$} & \colhead{$L_{cor}$} &
\colhead{$\eta$} & \colhead{$Q_{chrom}$} & \colhead{$Q_{cor}$} \\
\colhead{ } & \colhead{[s]} & \colhead{[$\rm s^{-1}$]} &
\colhead{[$\rm km/s$]} & \colhead{[Mm]} &
\colhead{ } & \colhead{[$\rm erg/s/cm^{3}$]} &
\colhead{[$\rm erg/s/cm^{3}$]} }
\startdata
21 &  60 & 0.04 & 1.48 & 49.6 & 0.346 & $6.08 \times 10^{-1}$ &
$2.97 \times 10^{-3}$ \\
22 & 100 & 0.04 & 1.48 & 49.6 & 0.399 & $6.30 \times 10^{-1}$ &
$2.30 \times 10^{-3}$ \\
23 & 200 & 0.04 & 1.48 & 49.6 & 0.440 & $5.03 \times 10^{-1}$ &
$1.89 \times 10^{-3}$ \\
24 &  60 & 0.02 & 0.74 & 49.6 & 0.434 & $1.29 \times 10^{-1}$ &
$0.90 \times 10^{-3}$ \\
25 &  60 & 0.06 & 2.21 & 49.6 & 0.350 & $1.59 \times 10^{0}$ &
$5.46 \times 10^{-3}$ \\
26 &  60 & 0.04 & 1.48 & 25.6 & 0.362 & $6.01 \times 10^{-1}$ &
$5.23 \times 10^{-3}$ \\
27 &  60 & 0.04 & 1.48 & 99.6 & 0.360 & $6.47 \times 10^{-1}$ &
$1.49 \times 10^{-3}$
\enddata
\end{deluxetable*}

\begin{figure*}
\epsscale{1.14}
\plotone{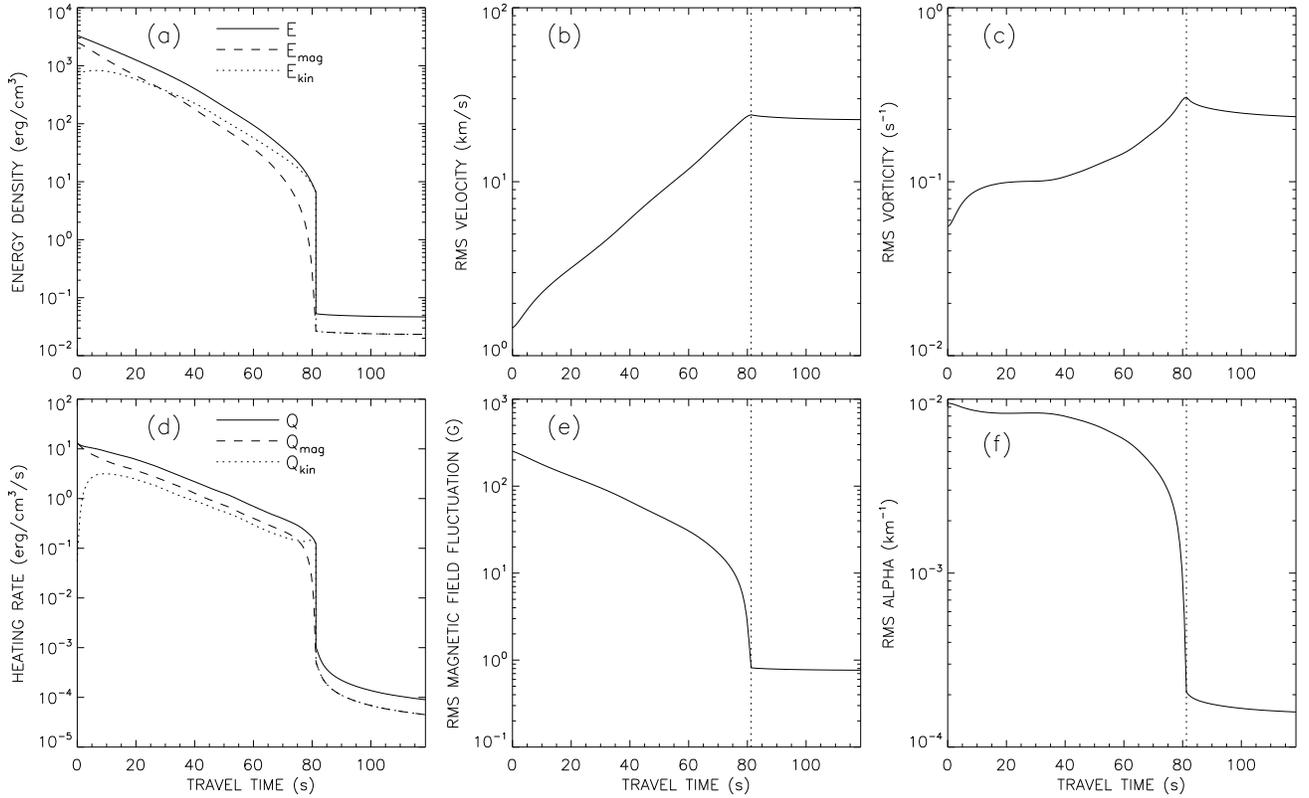}
\caption{Model for Alfv\'{e}n wave turbulence in an open field.
Various quantities are plotted as function of position along the
flux tube:
(a) kinetic and magnetic energy densities, and their sum $E(z)$;
(b) rms velocity $\Delta v_{rms}$; (c) rms vorticity $\omega_{rms}$;
(d) kinetic and magnetic heating rates, and their sum $Q(z)$;
(e) rms magnetic fluctuation $\Delta B_{rms}$; (f) rms twist
parameter $\alpha_{rms}$. Positions are given in terms of the Alfv\'{e}n
travel time $\tau (z)$. The corona is located in the range $81.3 <
\tau < 118.6$ s. The TR is indicated by the vertical {\it dotted}
line.}
\label{fig:open_figure3}
\end{figure*}
To demonstrate that the turbulence in the lower atmosphere is nearly
independent of processes in the corona, we construct an open field
model with the tube connected to the photosphere only at one end. In
this model there is only a single TR, and the coronal part of the tube
has constant temperature, density and field strength ($T_0 = 1.5$ MK,
$\rho_0 = 9 \times 10^{-15}$ $\rm g ~ cm^{-3}$, $B_{cor} = 50$ G). At
the coronal end of the tube we use open boundary conditions such that
the Alfv\'{e}n waves propagate out of the computational domain and no
waves are injected. In the coronal part of this model there are only
outward propagating waves, hence no nonlinear wave-wave interactions.
However, wave reflection still occurs in the chromosphere and TR, so
there are counter-propagating waves in the lower atmosphere, which
generates strong turbulence.  In Figure \ref{fig:open_figure3},
various quantities are plotted as function of position along the flux
tube for this open field model. Comparison of Figures
\ref{fig:figure3} and \ref{fig:open_figure3} indicates that the energy
density, heating rate and other parameters of the turbulence in the
lower atmosphere ($\tau < 81.3$ s) are very similar to those in the
closed field model. Figure \ref{fig:open_figure3}d shows the corona
(region with $\tau > 81.3$ s) is still being heated, but this residual
heating is due to the damping of the outward traveling waves, and is
not due to turbulence. 

In the above simplified model for the open field, there is no
turbulent decay of the Alfv\'{e}n waves in the corona. However, more
realistic models of coronal holes and other open-field structures
have shown that there is significant reflection of waves in the
corona, and that Alfv\'{e}n wave turbulence can explain the heating
and acceleration of the solar wind \citep[e.g.,][]{Dmitruk2002,
Cranmer2005, Cranmer2007, Chandran+Hollweg2009, Verdini2010}.

\subsection{Dependence of Heating Rate on Model Parameters}
\label{sect:Depend}

We construct a series of closed field models with different values of
the model parameters. Table \ref{table1} shows the subset of models in
which the properties of the footpoint motions or the coronal loop
length are varied. The first five columns in Table \ref{table1} show
the model name, the correlation time $\tau_0$, vorticity $\omega_0$
and velocity $\Delta v_{rms}$ of the footpoint motions, and the loop
length $L_{cor}$. Model 21 is the reference model described in the
previous subsection, and models 22 - 27 are variants in which one of
the parameters $\tau_0$, $\omega_0$ or $L_{cor}$ is changed relative
to the reference model. For all models listed in Table \ref{table1},
the coronal field strength $B_{cor} = 50$ G, the height of the
transition region $z_{TR} = 1.8$ Mm, the coronal pressure $p_{cor} =
1.833$ $\rm dyne ~ cm^{-2}$, the time step $\Delta t_0 = 0.746$ s,
the expansion factor $\Gamma = 1$, and damping rate $\nu_{max} = 0.7$
$\rm s^{-1}$.

For each model we simulate the dynamics of the Alfv\'{e}n waves
generated by random footpoint motions for a period of 3000 seconds,
and we compute the heating rate $Q(z)$ averaged over the time interval
$t = [800, 3000]$ seconds of the simulation. The period before
$t = 800$ [s] is omitted because it sometimes contains a large spike
in heating that may be an artifact of the initial conditions. The
heating rate $Q(z)$ decreases with height in the lower atmosphere.
The average heating rate in the chromosphere is determined as
follows. The function $Q(z)$ in the height range $z = [700,1300]$
km is fit with the following expression:
\begin{equation}
Q(z) \approx Q_{chrom} \left[ \frac{\rho_0 (z)}{\rho_{chrom}}
\right]^\eta , \label{eq:Qfit}
\end{equation}
where $\rho_{chrom}$ is the density at $z = 1000$ km ($\rho_{chrom}
\approx 3.6 \times 10^{-11}$ $\rm g ~ cm^{-3}$), and $Q_{chrom}$ and
$\eta$ are constants, which are determined by the fit. Therefore,
$Q_{chrom}$ is the average heating rate at height $z = 1000$ km. We
also measure the average coronal heating rate, $Q_{cor}$.
The values of $\eta$, $Q_{chrom}$ and $Q_{cor}$ are listed in the last
three columns of Table \ref{table1}. Based on the results in Table
\ref{table1}, the coronal heating rate can be approximated as
\begin{displaymath}
Q_{cor} \approx 2.97 \times 10^{-3}
\left( 0.45 + \frac{33} {\tau_0} \right)
\left( \frac{\omega_0} {\mbox{0.04 $\rm s^{-1}$}} \right)^{1.65}
\end{displaymath}
\begin{equation}
\times \,
\left( \frac{L_{cor}} {\mbox{50 Mm}} \right)^{-0.92}
~~ [{\rm erg ~ cm^{-3} ~ s^{-1} }] , \label{eq:Qcor1}
\end{equation}
where $\tau_0$ is in seconds. Therefore, the heating rate increases
nonlinearly with the vorticity $\omega_0$ of the footpoint motions,
and decreases approximately inversely with loop length $L_{cor}$.

Table \ref{table2} shows the subset of models in which the coronal
field strength and plasma pressure are varied. The pressure is
controlled by the TR height, $z_{TR}$ (see section
\ref{sect:Background}). The first six
columns in Table \ref{table2} show the model name, the coronal field
strength $B_{cor}$, the TR height $z_{TR}$, the coronal pressure
$p_{cor}$, the loop length $L_{cor}$, and the time step
$\Delta t_0$ used in the simulation. $L_{cor}$ is listed here because
it varies slightly between these models (see Appendix B). The values
of coronal field strength typically found in active regions lie in
the range 10 - 500 G. However, the present version of the RMHD code
has difficulty simulating Alfv\'{e}n waves in loops with $B_{cor} >
200$ G, which is due to the large Alfv\'{e}n speeds involved
($v_A > 7000$ $\rm km ~ s^{-1}$). Therefore, in the present paper we
only consider loops with $B_{cor}$ in the range 12 - 200 G.
For all models shown in Table \ref{table2}, the footpoint motions
are characterized by $\tau_0 = 60$ s, $\omega_0 = 0.04$ $\rm s^{-1}$,
and $\Delta v_{rms} = 1.48$ km/s. As before, the expansion factor
$\Gamma = 1$ and the damping rate $\nu_{max} = 0.7$ $\rm s^{-1}$.
For each model we compute the heating rate $Q(z)$ averaged over the
time interval $t = [800, 3000]$ seconds of the simulation, and we
derive the parameters $\eta$, $Q_{chrom}$ and $Q_{cor}$ (see last
three columns of Table \ref{table2}).

\begin{deluxetable*}{crccccccc}
\tablewidth{0pt}
\tablecaption{\label{table2}Dependence of Heating Rates on Coronal
Field Strength and Plasma Pressure}
\tablehead{
\colhead{Model} & \colhead{$B_{cor}$} & \colhead{$z_{TR}$} &
\colhead{$p_{cor}$} & \colhead{$L_{cor}$} & \colhead{$\Delta t_0$} &
\colhead{$\eta$} & \colhead{$Q_{chrom}$} & \colhead{$Q_{cor}$} \\
\colhead{ } & \colhead{[G]} & \colhead{[Mm]} &
\colhead{[$\rm dyne/cm^{2}$]} & \colhead{[Mm]} & \colhead{[s]} &
\colhead{ } & \colhead{[$\rm erg/s/cm^{3}$]} &
\colhead{[$\rm erg/s/cm^{3}$]} }
\startdata
54 & 200 & 1.5 & 4.585 & 50.3 & 0.296 & 0.544 & $1.268$
& $6.86 \times 10^{-3}$ \\
52 & 200 & 1.6 & 3.375 & 50.4 & 0.252 & 0.514 & $1.357$
& $6.76 \times 10^{-3}$ \\
53 & 200 & 1.7 & 2.487 & 50.6 & 0.216 & 0.496 & $1.386$
& $6.47 \times 10^{-3}$ \\
51 & 200 & 1.8 & 1.833 & 49.4 & 0.186 & 0.544 & $1.236$
& $5.43 \times 10^{-3}$ \\
55 & 150 & 1.5 & 4.585 & 50.0 & 0.392 & 0.501 & $1.062$
& $5.07 \times 10^{-3}$ \\
56 & 150 & 1.6 & 3.375 & 50.5 & 0.337 & 0.475 & $1.144$
& $5.12 \times 10^{-3}$ \\
57 & 150 & 1.7 & 2.487 & 49.4 & 0.289 & 0.489 & $1.099$
& $4.84 \times 10^{-3}$ \\
58 & 150 & 1.8 & 1.833 & 50.6 & 0.247 & 0.483 & $1.017$
& $4.61 \times 10^{-3}$ \\
38 & 100 & 1.6 & 3.375 & 50.6 & 0.507 & 0.427 & $7.89 \times 10^{-1}$
& $3.75 \times 10^{-3}$ \\
44 & 100 & 1.7 & 2.487 & 50.7 & 0.433 & 0.425 & $8.05 \times 10^{-1}$
& $3.76 \times 10^{-3}$ \\
29 & 100 & 1.8 & 1.833 & 49.5 & 0.373 & 0.436 & $8.19 \times 10^{-1}$
& $3.42 \times 10^{-3}$ \\
42 & 100 & 1.9 & 1.351 & 49.5 & 0.319 & 0.440 & $8.00 \times 10^{-1}$
& $3.36 \times 10^{-3}$ \\
40 & 100 & 2.0 & 0.996 & 50.5 & 0.274 & 0.431 & $8.06 \times 10^{-1}$
& $3.03 \times 10^{-3}$ \\
36 & 100 & 2.1 & 0.734 & 50.1 & 0.235 & 0.442 & $8.03 \times 10^{-1}$
& $2.66 \times 10^{-3}$ \\
30 &  50 & 1.6 & 3.375 & 49.3 & 1.016 & 0.405 & $5.63 \times 10^{-1}$
& $2.90 \times 10^{-3}$ \\
32 &  50 & 1.7 & 2.487 & 50.5 & 0.864 & 0.392 & $5.99 \times 10^{-1}$
& $2.93 \times 10^{-3}$ \\
21 &  50 & 1.8 & 1.833 & 49.6 & 0.746 & 0.346 & $6.08 \times 10^{-1}$
& $2.97 \times 10^{-3}$ \\
33 &  50 & 1.9 & 1.351 & 50.5 & 0.637 & 0.392 & $5.83 \times 10^{-1}$
& $2.71 \times 10^{-3}$ \\
34 &  50 & 2.0 & 0.996 & 50.6 & 0.549 & 0.360 & $6.03 \times 10^{-1}$
& $2.64 \times 10^{-3}$ \\
31 &  50 & 2.1 & 0.734 & 50.1 & 0.469 & 0.365 & $6.01 \times 10^{-1}$
& $2.35 \times 10^{-3}$ \\
35 &  50 & 2.3 & 0.399 & 50.4 & 0.347 & 0.371 & $5.75 \times 10^{-1}$
& $2.06 \times 10^{-3}$ \\
39 &  25 & 1.6 & 3.375 & 50.4 & 2.018 & 0.423 & $4.63 \times 10^{-1}$
& $2.23 \times 10^{-3}$ \\
45 &  25 & 1.7 & 2.487 & 50.3 & 1.721 & 0.403 & $4.44 \times 10^{-1}$
& $2.13 \times 10^{-3}$ \\
28 &  25 & 1.8 & 1.833 & 49.5 & 1.481 & 0.372 & $4.61 \times 10^{-1}$
& $2.05 \times 10^{-3}$ \\
43 &  25 & 1.9 & 1.351 & 50.3 & 1.269 & 0.390 & $4.46 \times 10^{-1}$
& $2.05 \times 10^{-3}$ \\
41 &  25 & 2.0 & 0.996 & 49.5 & 1.100 & 0.407 & $4.45 \times 10^{-1}$
& $1.88 \times 10^{-3}$ \\
37 &  25 & 2.1 & 0.734 & 50.4 & 0.944 & 0.422 & $4.19 \times 10^{-1}$
& $1.85 \times 10^{-3}$ \\
46 &  25 & 2.3 & 0.399 & 50.4 & 0.694 & 0.389 & $4.31 \times 10^{-1}$
& $1.61 \times 10^{-3}$ \\
50 &  12 & 1.9 & 1.351 & 49.6 & 2.669 & 0.414 & $3.51 \times 10^{-1}$
& $1.39 \times 10^{-3}$ \\
49 &  12 & 2.0 & 0.996 & 50.5 & 2.283 & 0.406 & $3.70 \times 10^{-1}$
& $1.33 \times 10^{-3}$ \\
48 &  12 & 2.1 & 0.734 & 50.4 & 1.967 & 0.382 & $3.61 \times 10^{-1}$
& $1.30 \times 10^{-3}$ \\
47 &  12 & 2.3 & 0.399 & 50.1 & 1.438 & 0.373 & $3.52 \times 10^{-1}$
& $1.21 \times 10^{-3}$
\enddata
\end{deluxetable*}

\begin{figure*}
\epsscale{1.05}
\plotone{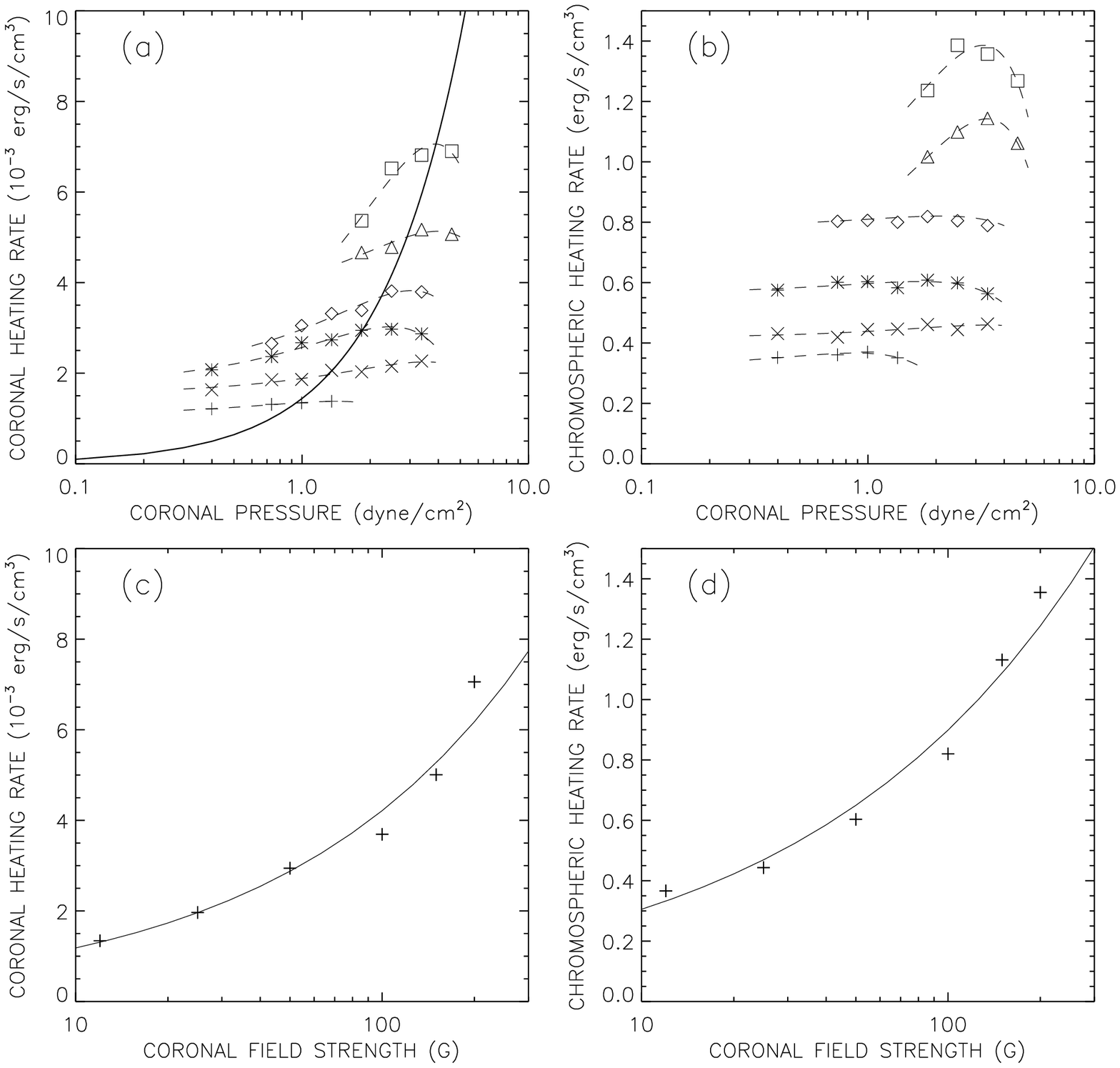}
\caption{%
Dependence of coronal and chromospheric heating rates on coronal
pressure ($p_{cor}$) and field strength ($B_{cor}$) for a loop with
constant cross-section and length $L_{cor} = 50$ Mm. (a) Average
coronal heating rate $Q_{cor}$ as function of $p_{cor}$. The symbols
show results from the numerical models listed in Table \ref{table2}
for $B_{cor} = 12$ G ({\it plusses}), 25 G ({\it crosses}), 50
({\it stars}), 100 G ({\it diamonds}), 150 G ({\it triangles}),
200 G ({\it squares}).
The {\it dashed} curves show quadratic fits to these data, and the
{\it solid} curve shows the heating rate predicted by the RTV scaling
law, equation (\ref{eq:QcorRTV}). The latter represents the condition
of thermal equilibrium (balance between heating and cooling
processes). (b) Heating rate $Q_{chrom}$ at height $z = 1$ Mm in the
chromosphere, plotted as function of coronal pressure. Panels (c)
and (d) show the equilibrium heating rates $Q_{cor}$ and $Q_{chrom}$
as function of coronal field strength. The {\it plus} signs show
the values derived from the intersections of the dashed and solid
curves in panel (a), and the {\it solid} curves show power-law fits
to these data, see equations (\ref{eq:Qcor2}) and (\ref{eq:Qchrom2}).}
\label{fig:Qcor}
\end{figure*}
Figures \ref{fig:Qcor}a and \ref{fig:Qcor}b show the coronal and
chromospheric heating rates as function of coronal pressure $p_{cor}$
for six values of coronal field strength ($B_{cor}$ = 12, 25, 50,
100, 150 and 200 G). The symbols show the simulation data listed in
Table \ref{table2}, and the dashed curves show quadratic fits to these
data. Both $Q_{cor}$ and $Q_{chrom}$ increase with coronal field
strength. The decrease of $Q_{cor}$ for low coronal pressures may be
explained as effects of wave reflection: for small $p_{cor}$ or high
$B_{cor}$ the coronal Alfv\'{e}n speed is very high, resulting in
strong wave reflection in the chromosphere and TR. As a result, a
larger fraction of the wave energy is dissipated in the lower
atmosphere. Figure \ref{fig:Qcor}b shows that the chromospheric
heating rate depends only weakly on coronal pressure.

As discussed in section \ref{sect:Background}, the peak temperature
$T_{max}$ along a coronal loop was chosen to be consistent with the
first RTV scaling law, equation (\ref{eq:Tmax}), but the heating rates
$Q_{cor}$ found in the simulations are not necessarily consistent with
the second scaling law, equation (\ref{eq:QcorRTV}). Therefore, the
models listed in Tables \ref{table1} and \ref{table2} are not
necessarily in {\it thermal} equilibrium. The solid curve in Figure
\ref{fig:Qcor}a shows the heating rate $Q_{cor}$ derived from equation
(\ref{eq:QcorRTV}) with $L_{cor} = 50$ Mm. This curve shows the value
of $Q_{cor}$ at which the coronal loop is in thermal equilibrium
(i.e., the heating is balanced by radiative and conductive losses).
To the left of this curve, the time-averaged heating rate due to
Alfv\'{e}n wave turbulence is larger than the radiative and conductive
losses, so the coronal temperature will increase, more mass will be
evaporated into the corona from the chromosphere, and the coronal
pressure will increase. The opposite happens to the right of the solid
curve. Therefore, although the present models do not include the
effects of chromospheric evaporation, it is clear that in models with
time-dependent coronal pressure the corona will have a tendency to
approach the equilibrium state represented by the RTV scaling laws.

The field-strength dependence of the equilibrium heating rate
$Q_{cor}$ can be determined by finding the intersection points of the
solid and dashed curves in Figure \ref{fig:Qcor}a, and the
corresponding chromospheric heating rate $Q_{chrom}$ can be obtained
from Figure \ref{fig:Qcor}b. The ``equilibrium'' heating rates are
plotted in Figures \ref{fig:Qcor}c and \ref{fig:Qcor}d as function of
coronal field strength, $B_{cor}$. Note that both heating rates
increase with coronal field strength, and show no sign of saturation
for large $B_{cor}$. The dependences on $B_{cor}$ can be roughly
approximated as power laws:
\begin{eqnarray}
Q_{cor} & \approx & 2.88 \times 10^{-3} \left( \frac{B_{cor}}
{\mbox{50 G}} \right)^{0.55} ~~~ [{\rm erg ~ cm^{-3} ~ s^{-1} }] ,
\label{eq:Qcor2} \\
Q_{chrom} & \approx & 6.49 \times 10^{-1} \left( \frac{B_{cor}}
{\mbox{50 G}} \right)^{0.47} ~~~ [{\rm erg ~ cm^{-3} ~ s^{-1} }] ,
\label{eq:Qchrom2} 
\end{eqnarray}
although the fits are not very accurate at high field strength (see
{\it solid} curves in Figures \ref{fig:Qcor}c and \ref{fig:Qcor}d).
Combining equations (\ref{eq:Qcor1}) and (\ref{eq:Qcor2}), we obtain
\begin{displaymath}
Q_{cor} \approx 2.9 \times 10^{-3}
\left( 0.45 + \frac{33} {\tau_0} \right)
\left( \frac{\Delta v_{rms}} {\mbox{1.48 km/s}} \right)^{1.65}
\end{displaymath}
\begin{equation}
\times \,
\left( \frac{L_{cor}} {\mbox{50 Mm}} \right)^{-0.92}
\left( \frac{B_{cor}} {\mbox{50 G}} \right)^{0.55} 
~ [{\rm erg ~ cm^{-3} ~ s^{-1} }] , \label{eq:Qcor3}
\end{equation}
where we replaced the vorticity $\omega_0$ by the velocity $\Delta
v_{rms}$ of the footpoint motions. Note that the equilibrium heating
rate depends on coronal field strength $B_{cor}$, on loop length
$L_{cor}$, and on the parameters of the footpoint motions ($\tau_0$
and $\Delta v_{rms}$). 

The above expression assumes that the coronal temperature and pressure
have the values predicted by the RTV scaling laws, and therefore
neglects the effects of the spatial and temporal variability of the
heating \citep[][]{Klimchuk2006}.  Also, it should be kept in mind
that the numerical simulations have been done only in a limited range
of coronal field strength and loop length ($12 < B_{cor} < 200$ G, $25
< L_{cor} < 100$ Mm), so the above expression for $Q_{cor}$ is valid
only within this limited range.  Furthermore, the above expression
applies only to coronal loops with constant cross-section ($\Gamma =
1$), and the effects of gravity have been neglected in the corona (but
not in the lower atmosphere).  Expression (\ref{eq:Qcor3}) should not
be applied to coronal loops on the quiet Sun, where magnetic fields
are weaker and the effects of gravity and coronal loop expansion
cannot be neglected.

We also considered how the heating rates depend on the maximum
damping rate $\nu_{max}$ of the Alfv\'{e}n modes [see equation
(\ref{eq:nuk})].  For model Nos.~21, 28 and 29 we did a second
simulation with twice the damping rate ($\nu_{max} = 1.4$
$\rm s^{-1}$), and we found that the coronal heating rates $Q_{cor}$
are changed by about +1\%, -6\% and -6\%, respectively. The
chromospheric heating rates $Q_{chrom}$ are changed by -6\%, +2\% and
-10\% for these three models. These numbers are relatively small
compared to the factor 2 change in the damping rate, indicating that
the rate of turbulent dissipation is insensitive to the value of the
damping rate of the high wavenumber modes, as one would expect for a
turbulent process.

\subsection{Effects of Coronal Loop Expansion}
\label{sect:Expansion}

\begin{deluxetable*}{ccccccccc}
\tablewidth{0pt}
\tablecaption{\label{table3}Effect of Coronal Loop Expansion}
\tablehead{
\colhead{$\Gamma$} & \colhead{$V_{cor}$} & \colhead{$\eta$} &
\colhead{$Q_{chrom}$} & \colhead{$Q_{cor}$} & \colhead{$F_{A} (z_{TR})$}
& \colhead{$P_{tot}$} & \colhead{$P_{cor}$} & \colhead{$f_{cor}$} \\
\colhead{ } & \colhead{[$\rm cm^{-3}$]} & \colhead{ } &
\colhead{[$\rm erg/s/cm^{3}$]} &
\colhead{[$\rm erg/s/cm^{3}$]} &
\colhead{[$\rm erg/s/cm^{2}$]} &
\colhead{[$\rm erg/s$]} & \colhead{[$\rm erg/s$]} & \colhead{ } }
\startdata
1 & $0.44 \times 10^{26}$ & 0.346 & $6.08 \times 10^{-1}$ &
$2.97 \times 10^{-3}$ & $0.77 \times 10^7$ & $1.69 \times 10^{24}$ &
$1.30 \times 10^{23}$ & 0.077 \\
2 & $0.73 \times 10^{26}$ & 0.386 & $5.70 \times 10^{-1}$ &
$2.54 \times 10^{-3}$ & $1.09 \times 10^7$ & $1.66 \times 10^{24}$ &
$1.85 \times 10^{23}$ & 0.111 \\
3 & $1.03 \times 10^{26}$ & 0.353 & $5.57 \times 10^{-1}$ &
$2.11 \times 10^{-3}$ & $1.25 \times 10^7$ & $1.66 \times 10^{24}$ &
$2.17 \times 10^{23}$ & 0.130 \\
4 & $1.32 \times 10^{26}$ & 0.377 & $5.75 \times 10^{-1}$ &
$1.76 \times 10^{-3}$ & $1.34 \times 10^7$ & $1.70 \times 10^{24}$ &
$2.32 \times 10^{23}$ & 0.137 \\
6 & $1.91 \times 10^{26}$ & 0.371 & $5.53 \times 10^{-1}$ &
$1.43 \times 10^{-3}$ & $1.56 \times 10^7$ & $1.69 \times 10^{24}$ &
$2.73 \times 10^{23}$ & 0.161
\enddata
\end{deluxetable*}

Models of the coronal magnetic field \citep[e.g.,][]{Schrijver2005}
predict that the field strength $B_0$ generally decreases with height
in the corona. Therefore, we expect that the cross-sectional area $A$
of a coronal loop increases with height: $A = \Phi/B_0$, where $\Phi$
is the magnetic flux, which presumably is constant along the loop.
However, observed X-ray and EUV loops often show approximately
constant cross-section \citep[][]{Klimchuk2000, Watko2000,
LopezFuentes2006}. The reasons for this discrepancy are not well
understood. The present paper cannot directly
address this issue because our model does not include interactions
between neighboring flux tubes, which we believe is important for
resolving this issue. However, we can investigate how Alfv\'{e}n waves
are affected by the expansion of the loop with height. We define the
areal expansion factor $\Gamma$ as described in equation
(\ref{eq:B0}). This factor affects only the {\it coronal} part of the
loop, and has no effect on the magnetic field $B_0(z)$ in the lower
atmosphere. We construct a series of models with different expansion
factors, but otherwise identical model parameters ($z_{TR} = 1.8$ Mm,
$L_{cor} = 50$ Mm, $B_{cor} = 50$ G). For each model we compute the
heating rate per unit volume, $Q(z)$, averaged over the time interval
from 800 to 3000 seconds and over the cross-sectional area of the flux
tube. We also compute the rate of energy input into the corona,
\begin{equation}
P_{cor}^\prime = A(z_{TR1}) F_A(z_{TR1}) - A(z_{TR2}) F_A(z_{TR2}) ,
\end{equation}
where $A(z) = \pi R^2(z) $ is the cross-sectional area of the
tube, $z_{TR1}$ and $z_{TR2}$ are the positions of the TRs, and
$F_A (z)$ is the time-averaged Alfv\'{e}n wave energy flux [see equation
(\ref{eq:flux})]. Finally, we compute two energy dissipation rates,
$P_{tot}$ and $P_{cor}$, which represent the dissipated power
integrated over the total volume of the tube and over the coronal
volume, respectively:
\begin{equation}
P_{tot} \equiv \int_0^L Q(z) A(z) dz , ~~~~
P_{cor} \equiv \int_{z_{TR1}}^{z_{TR2}} Q(z) A(z) dz .
\end{equation}
We find that $P_{cor}$ and $P_{cor}^\prime$ are equal to within a few
percent, as expected for a statistically stationary state.
This equality also shows that energy is very well conserved in these
numerical models.

The results of the simulations are summarized in Table \ref{table3}.
The nine columns of this table show the areal expansion factor
$\Gamma$, the coronal volume $V_{cor}$, the parameters $\eta$ and
$Q_{chrom}$ describing the heating in the lower atmosphere [see
equation (\ref{eq:Qfit})], the volume-averaged coronal heating rate
$Q_{cor}$, the Alfv\'{e}n wave energy flux $F_A (z_{TR})$ at the TR, the
integrated dissipation rates $P_{tot}$ and $P_{cor}$, and the fraction
of energy dissipated in the corona, $f_{cor} \equiv P_{cor} /
P_{tot}$. The model with $\Gamma = 1$ is the reference model discussed
in section \ref{sect:RefMod}.
We find that for the reference model only 7.7\% of the
available energy is dissipated in the corona; the remainder is
dissipated in the photosphere and chromosphere at the two ends of the
coronal loop. The energy flux $F_A (z_{TR})$ increases with
the expansion factor, and so does the coronal power $P_{cor}$ and the
fraction $f_{cor}$. Specifically, for $\Gamma = 6$ about 16\% of the
available energy is dissipated in the corona. In contrast, the total
dissipation rate $P_{tot}$ is almost independent of the expansion
factor. Therefore, the heating of the lower atmosphere is apparently
not sensitive to the properties of the coronal part of the loop.

The results of Table \ref{table3} show that, as $\Gamma$ increases,
more energy enters into the corona and is dissipated at coronal
heights. We suggest this increase is due to the fact that for large
$\Gamma$ the Alfv\'{e}n speed at the midpoint of the coronal loop ($z =
L/2$) is drastically reduced compared to the case $\Gamma = 1$.
As a result, it takes a longer time for the Alfv\'{e}n waves to travel
through the corona, and this gives the waves more time to dissipate
their energy in the corona via turbulent processes. The wave energy
is distributed over a coronal volume $V_{cor}$ that increases more
rapidly with $\Gamma$ than the energy dissipation rate. Therefore,
the average heating rate $Q_{cor}$ in the corona decreases with
increasing $\Gamma$ (see Table \ref{table3}).

\begin{figure*}
\epsscale{1.00}
\plotone{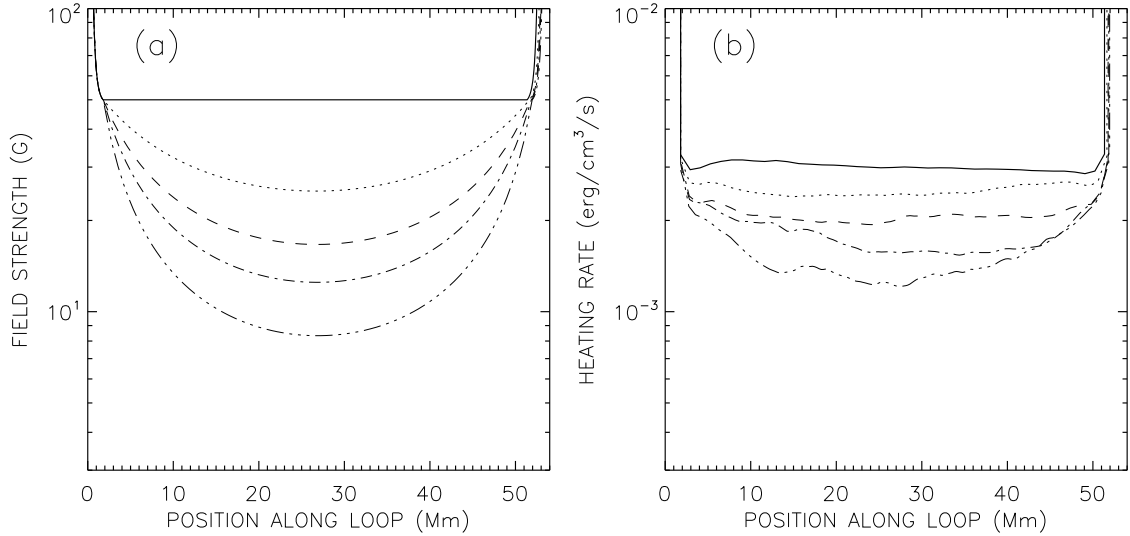}
\caption{Effects of coronal loop expansion. (a) Magnetic field
strength $B_0 (z)$, and (b) volumetric heating rate $Q(z)$, as
functions of position $z$ along the coronal loop for different values
of the loop expansion factor:
$\Gamma = 1$ ({\it full}), $\Gamma = 2$ ({\it dotted}),
$\Gamma = 3$ ({\it dashed}), $\Gamma = 4$ ({\it dash-dotted}), and
$\Gamma = 6$ ({\it dash-triple-dotted}).}
\label{fig:expand}
\end{figure*}
Figure \ref{fig:expand} shows the magnetic field strength $B_0 (z)$
and time-averaged volumetric heating rate $Q(z)$ as functions of
position along the
coronal loop. Note that $Q(z)$ is relatively constant within the
corona, and varies more gradually than the magnetic field strength
$B_0(z)$. A power-law fit to these data for $\Gamma =$ 4 and 6 yields
$Q(z) \propto [B_0 (z)]^{0.4}$, so the exponent is significantly
less than unity. We suggest this relative constancy of the volumetric
heating rate may be due to Alfv\'{e}n wave reflection within the
corona. For large $\Gamma$, the Alfv\'{e}n speed $v_A (z)$ has a local
minimum at the mid-point of the loop ($z = 27$ Mm), and increases
toward the ends. Hence, as the Alfv\'{e}n waves travel from the
midpoint to the ends of the loop, the waves are partially reflected
before they reach the TR. This enhances the wave energy density in the
central part of the loop, and gives the wave more time to be
dissipated near the midpoint. Clearly, wave reflections in the
chromosphere, at the TR, and within the corona play an important role
in determining the spatial distribution of the average heating rate
$Q(z)$.

\begin{figure}
\epsscale{1.13}
\plotone{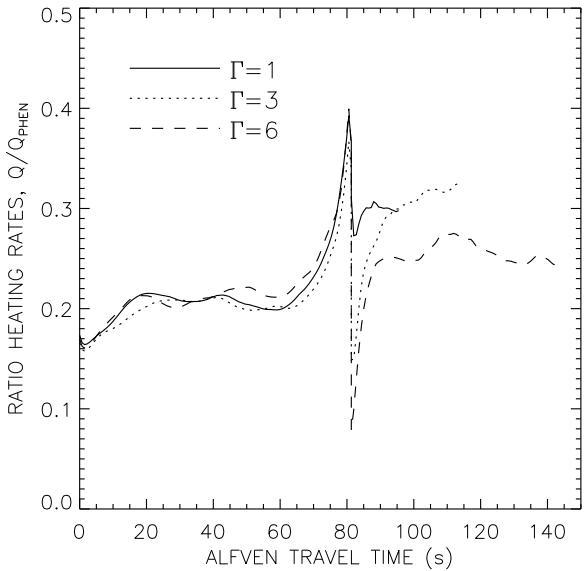}
\caption{Ratio of the numerically computed heating rate $Q(z)$ and the
rate $Q_{phen}(z)$ predicted by phenomenological turbulence models.
The ratio $q \equiv Q/Q_{phen}$ is plotted as function of Alfv\'{e}n
travel time $\tau(z)$ from the left footpoint for three different
values of the loop expansion factor: $\Gamma = 1$ ({\it full}),
$\Gamma = 3$ ({\it dotted}), $\Gamma = 6$ ({\it dashed}). Only half of
the loop is shown ($0 < z < L/2$).}
\label{fig:ratio}
\end{figure}
It is worthwhile to compare the computed heating rates $Q(z)$ with
those predicted from phenomenological turbulence models \citep[][]
{Hossain1995, Matthaeus1999, Dmitruk2001, Dmitruk2002, Chandran2009}.
In such models the fluctuations are expressed in terms of the
Elsasser variables, here defined as ${\bf Z}_\pm \equiv {\bf v}
\pm {\bf B}_\perp / \sqrt{4 \pi \rho_0}$. The turbulent dissipation
rate is given by equation (57) of \citet{Cranmer2005}:
\begin{equation}
Q_{phen} (z) = \rho_0 \frac{Z_-^2 Z_+ + Z_-^2 Z_+} {4 L_\perp} ,
\end{equation}
where $L_\perp (z) $ is the outer scale of the turbulence, and the
quantities $Z_\pm (z)$ are the rms values of the Elsasser
variables. The latter are averaged over the cross-section of the flux
tube [see equation (\ref{eq:DZrms})] and over time. We compute
$Z_\pm (z)$ for several models and find that the turbulence is
significantly unbalanced ($Z_+ \ne Z_-$), which is due to the strong
dissipation in our models. The perpendicular length scale is set equal
to the flux tube radius, $L_\perp = R(z)$. Figure \ref{fig:ratio}
shows the ratio $q(z) \equiv Q(z) / Q_{phen}(z)$ plotted as function
of position along the flux tube for three values of the coronal loop
expansion factor ($\Gamma = 1$, 3 and 6). Positions are given in terms
of the Alfv\'{e}n travel time $\tau (z)$ from the left footpoint. We
find that $q \approx 0.2$ in the photosphere and low chromosphere, and
the ratio peaks in the upper chromosphere, indicating the turbulence
is more efficient there. For the reference model $q \approx 0.3$ in
the corona ({\it solid} curve), similar to the value of 0.5 used by
\citet{Breech2009} in their model of the solar wind. Similar values
for $q$ are found for the models with larger expansion factors (the
downward spikes just above the TR may be due to insufficient spatial
resolution). We conclude that $q$ is significantly less than unity at
all heights.

\section{Summary and Discussion}
\label{sect:Discussion}

According to the present model, the interactions of turbulent
convective flows with kilogauss flux tubes in the photosphere produce
transverse displacements of magnetic field lines on length scales less
than the width of the flux tubes, i.e., less than about $100$ km
(see Figure \ref{fig:cartoon}). We assumed that the photospheric
footpoint motions have velocity amplitudes in the range 1 - 2 km/s and
correlation times of 60 - 200 s. We studied the dynamics of the
Alfv\'{e}n waves produced by such footpoint motions, and found that
such motions produce Alfv\'{e}nic turbulence at larger heights in
the flux tube. The predicted dissipation rates of the turbulence as
function of height are sufficient to reproduce the observed rates of
chromospheric and coronal heating in active regions. We conclude that
{\it fine scale magnetic braiding can drive Alfv\'{e}nic turbulence
and can produce sufficient heating for both the chromosphere and
corona}. However, we should emphasize that we do not have any direct
observational evidence for the existence of the assumed footpoint
motions, as scales of $100$ km or less cannot be resolved with
present-day solar telescopes.  Therefore, our conclusion can be made
only for the chosen values of the photospheric driver; for higher or
lower values the heating does not meet the observational constraints.
Our results indirectly provide a constraint on the amplitude of the
footpoint motions necessary for the proposed mechanism to heat the
chromosphere and corona. Our model produces a coronal heating profile
that is similar to that of a nanoflare storm \citep[][]{Klimchuk2006}
in the spatial and temporal distribution of heating events, and in the
final source of the energy released - dissipation of magnetic fields.
The difference is that we have provided a fully self-consistent source
for the nanoflare energy. The source depends critically on the
existence of a chromospheric reservoir of turbulent Alfv\'{e}n waves.

To investigate the nonlinear dynamics of the Alfv\'{e}n waves, we
considered a flux tube that extends from the photosphere through the
chromosphere into the corona.  We mainly considered closed coronal
loops, and for simplicity we assumed that a single kilogauss flux tube
at one end of the loop is connected to a single flux tube at the other
end. Therefore, we ignored the fact that on the real Sun the
photospheric flux concentrations at the two ends are uncorrelated and
do not perfectly match up. Furthermore, we assumed that the flux
elements do not split up or merge during the simulation, so that the
flux tube retains its identity. The pressure, density and magnetic
field strength were taken to be fixed functions of position $z$ along
the flux tube. The assumed structure of the lower atmosphere is based
on model P by \citet{Fontenla2009}; this model represents faculae in
active regions (i.e., very bright plage). The chromosphere-corona TR
was treated as a discontinuity where waves can reflect, and we
neglected the gravitational stratification in the corona, so the
pressure is constant along the coronal part of the loop.

The dynamics of the waves and their structure in the plane
perpendicular to the loop axis were described using the RMHD
approximation \citep[][]{Strauss1976}, which assumes that the magnetic
and velocity perturbations of the waves are perpendicular to the mean
magnetic field of the flux tube. We found that the Alfv\'{e}n waves
strongly reflect as they propagate through the chromosphere. This
reflection is due to the increase of Alfv\'{e}n speed with
height. Even stronger reflection occurs at the TR where the Alfv\'{e}n
speed suddenly increases by about a factor of 15. These reflections
produce in the chromosphere a pattern of counter-propagating waves
that are subject to nonlinear wave-wave interactions. We found that
the waves quickly decay into turbulence, causing most of the wave energy
to be deposited in the lower atmosphere (photosphere and
chromosphere). Only a small fraction of the wave energy is transmitted
into the corona. Energy is injected at both ends of the coronal loop,
producing in the corona two sets of counter-propagating waves that
reflect off the TRs and decay into turbulence. The dissipation is
entirely due to Alfv\'{e}n wave turbulence, and is not due to phase
mixing or resonant absorption \citep[e.g.,][]{Heyvaerts1983,
DeGroof2002}, which are not included in the present model.

We constructed a variety of loop models with different values of the
model parameters. For the reference model (section \ref{sect:RefMod})
we assumed the photospheric footpoint motions have an rms velocity of
1.48 km/s and a correlation time $\tau_0 = 60$ s. We found that
the heating rate $Q(z,t)$ varies strongly with position $z$ along the
loop and with time, which is due to bursts of wave activity and
subsequent dissipation of the waves via turbulence. 
Only a small fraction of the
wave energy is dissipated in the corona (about 7.7\%), and the
remainder is dissipated in the lower atmospheres at the two ends of
the modeled loop. We computed the time-averaged
heating rate $Q(z)$ and found that this rate decreases with height in
the lower atmosphere from about 10 $\rm erg ~ cm^{-3} ~ s^{-1}$ in the
photosphere to about 0.2 $\rm erg ~ cm^{-3} ~ s^{-1}$ in the upper
chromosphere. These values can be compared with rates of radiative
loss found in semiempirical models of the solar atmosphere
\citep[e.g.,][]{Avrett1985, Anderson1989, Fontenla1999, Fontenla2006,
Fontenla2009}. \citet{Fontenla2009} computed radiative loss rates as
function of gas pressure (i.e., height) for several different
wavelength bands (see their Figure~14). For model P the largest
radiative loss occurs in the 2000-3000 {\AA} band, which contains the
Mg~II resonance lines. The total losses over all wavelengths are in
the range 1 - 2 $\rm erg ~ cm^{-3} ~ s^{-1}$, somewhat larger than the
values found in the present model. Therefore, our model may
underestimate the amplitude of the Alfv\'{e}n waves and rate of energy
dissipation in the chromosphere.

In the corona the reference model predicts $Q_{cor} \sim 3 \times
10^{-3}$ $\rm erg ~ cm^{-3} ~ s^{-1}$, which is sufficient to explain
the heating of typical active region loops. The predicted velocity
amplitude of the Alfv\'{e}n waves in the corona is in the range
20 - 40 km/s, similar to the values found in observations of spectral
line widths in active regions \citep[][]{Dere1993, Warren2008,
Li2009}. We investigated how the coronal and chromospheric heating
rates depend on the model parameters (see section \ref{sect:Depend}).
We used the RTV scaling laws to determine the ``equilibrium'' heating
rates for which the coronal heating is balanced by radiative and
conductive losses, and derived a scaling law describing how the
coronal heating rate $Q_{cor}$ depends on the coronal field strength,
loop length, and the parameters of the footpoint motions [equation
(\ref{eq:Qcor3})].  This result may be compared with other models of
coronal heating \citep[see][]{Mandrini2000, Schrijver2004}.

We also considered coronal loops with non-constant cross-section
(section \ref{sect:Expansion}) and found that both the energy flux
$F_A (z_{TR})$ entering the corona and the total power $P_{cor}$
dissipated in the corona increases with the areal expansion factor
$\Gamma$. The increase of $P_{cor}$ with $\Gamma$ is due to the
decrease in the coronal Alfv\'{e}n speed, which lengthens the
Alfv\'{e}n travel time in the corona and gives the waves more time to
dissipate their energy via turbulence. The fraction of energy
dissipated in the corona also increases with $\Gamma$, but is still
only about 16\% for the model with $\Gamma = 6$. The average coronal
heating rate $Q_{cor}$ decreases with $\Gamma$, which is due to the
fact that the coronal volume increases faster than the dissipated
power $P_{cor}$. We found that for models with $\Gamma > 1$ the
heating rate $Q(z)$ varies with position along the coronal part of the
loop, and decreases towards the midpoint of the loop where the field
strength $B_0(z)$ is lowest.  However, $Q(z)$ varies less than
$B_0(z)$, so the heating is not strongly concentrated near the coronal
base. This effect can be understood in terms of reflection of the
waves by gradients of Alfv\'{e}n speed in the corona.

Observations of the active corona using the Transition Region and
Coronal Explorer \citep[TRACE, see][]{Handy1999} have shown that
the corona is highly dynamic and full of flows and wave phenomena
\citep[e.g.,][]{Schrijver1999}. This has led some authors to suggest
that the heating of active region loops is localized in the low
corona and involves energization processes that operate in the
chromosphere and transition region \citep[][]{Aschwanden2001,
Aschwanden2007, DePontieu2009}. However, \citet{Klimchuk2010} show
that heating concentrated in the low corona would lead to thermal
non-equilibrium of the plasma, and such non-equilibrium is not
consistent with the observed properties of warm (1-2 MK) loops
observed in active regions. Instead, they suggest coronal loops are
heated impulsively by storms of nanoflares that are distributed
throughout the coronal loop \citep[also see][]{Parker1988, Lu1991,
Klimchuk2006, Klimchuk2009}. The present model has some features
in common with both types of models. On the one hand, we find that the
average heating in the chromosphere is 2 or 3 orders of magnitude
larger than that in the low corona, and that most of the heating
occurs in the lower atmosphere, which is consistent with the ideas of
\citet{DePontieu2009}. On the other hand, in the coronal part of the
loop the heating depends only weakly on position, so the loop is
likely to be thermally stable, consistent with the results of
\citet{Klimchuk2010}.

Although the present model was described in terms of the propagation
and dissipation of Alfv\'{e}n waves, the waves reflect at various
heights and undergo nonlinear interactions, producing twisted and
braided fields similar to those found in previous braiding models
\citep[e.g.,][]{Mikic1989, Longcope1994, Hendrix1996b, Galsgaard1996,
Ng1998, Craig2005, Rappazzo2008}. Also, the model includes the effects
of magnetic reconnection, and the time-dependent heating seen in
Figure \ref{fig:figure2}d can be thought of as a sequence of
``nanoflares'' \citep[][]{Parker1988}. However, unlike in previous
``quasi-static'' braiding models the twisted structures are highly
dynamic and are far from the force-free equilibrium state.  For
example, the twist parameter $\alpha$ is not at all constant along the
field lines (see Figure \ref{fig:figure3}f), as would be expected for
a force-free state. This is due to the inclusion of the lower
atmospheres at the two ends of the coronal loop, which produces
turbulence in the chromosphere. Also, the transverse motions of field
lines at the photospheric base occur entirely {\it inside} the
magnetic flux elements, hence these motions do not contribute to the
``random walk'' of the flux elements on the photosphere. This has the
advantage that the velocity amplitudes can be relatively large, yet be
consistent with observational constraints on the photospheric
diffusion constant \citep[][] {DeVore1985, Berger1998}.

According to the present model, the transverse scale of the braids in
the corona is very small, $\ell_\perp < R_{cor} = 529$ km, so the
braids would likely not be observable with existing X-ray or EUV
telescopes. Furthermore, the rms value of the transverse field
fluctuations is small compared to the mean field, $\Delta B_{rms} /B_0
\approx 0.025$, so the mis-alignment angle between the field lines and
the tube axis is only about $1.4^\circ$.  Both the transverse scale of
the braids and the mis-alignment angle are consistent with the fact
that no braiding is observed in the corona on scales of several
arcseconds or larger (see section \ref{sect:Obs}). The mis-alignment
angle is much smaller than that required in quasi-static braiding
models that match the observed heating rate \citep[][]{Parker1983,
Priest2002}. We conclude that magnetic energy is injected into the
corona via a dynamic (not quasi-static) braiding process. The
existence of turbulence in coronal loops may have important
implications for the damping of coronal loop oscillations \citep[][]
{Nakariakov1999, Ofman2002, Roberts2008} and for the cross-field
diffusion of electrons \citep[][]{Galloway2006}.

In the present paper we did not consider the plasma response to the
heating. In future work we will construct models of the thermal
structure of coronal loops heated by Alfv\'{e}n wave turbulence,
including the effects of thermal conduction and radiative losses. 
The results will be compared with predictions from quasi-static
braiding models. Another line of future research will be to investigate
how the temporal fluctuations of the heating rate affect the
temperature and density. Is thermal conduction strong enough to smooth
out the spatial and temporal fluctuations? Ideally, such studies
should take into account the non-uniformity of the temperature and
density in the perpendicular plane; this may require a more complete
3D MHD model. We also intend to consider the interactions between
neighboring flux tubes in the lower atmosphere, including the effects
of the splitting and merging of kilogauss flux tubes.

\acknowledgements
We thank the anonymous referee for many useful comments which helped
to improve this paper. We also thank B.~Chandran for discussions on
nonlinear effects in Alfv\'{e}nic turbulence.
Hinode is a Japanese mission developed and launched by ISAS/JAXA,
with NAOJ as domestic partner and NASA and STFC (UK) as international
partners. It is operated by these agencies in cooperation with ESA
and NSC (Norway). This work was supported by NASA contract NNM07AB07C
(NASA Solar-B X-Ray Telescope Phase-E). This research has made use of
NASA's Astrophysics Data System.

\appendix

\section{A. Comparison With Transport Equations Based on Elsasser
Variables}
\label{sect:Elsasser}

In section \ref{sect:RMHD} we formulated the Alfv\'{e}n wave transport
equations in terms of scalar fields $f$, $h$, $\omega$ and $\alpha$.
In contrast, previous works often used the Elsasser variables
\citep[][]{Elsasser1950}, which are vector fields:
\begin{equation}
{\bf Z}_\pm ({\bf r},t) \equiv {\bf v}_\perp \pm \frac{{\bf B}_\perp}
{\sqrt{4 \pi \rho_0}} ,
\end{equation}
where ${\bf v}_\perp = \nabla f \times \hat{\bf B}_0$ is the
velocity, and ${\bf B}_\perp = \nabla h \times {\bf B}_0$ is the
perturbation of the magnetic field. To demonstrate that the present
formulation of the transport equations is equivalent to that used
by others, we start with equation (21) of \citet{Zhou1990}:
\begin{eqnarray}
\frac{\partial {\bf Z}_\pm} {\partial t} & = &
- \left( {\bf U} \mp {\bf V}_A \right) \cdot \nabla {\bf Z}_\pm 
- \onehalf \left( {\bf Z}_\pm  - {\bf Z}_\mp \right) \nabla \cdot
\left( \onehalf {\bf U} \pm {\bf V}_A \right) \nonumber \\  
 &  & - {\bf Z}_\mp \cdot \left( \nabla {\bf U} \pm \frac{1}
{\sqrt{4 \pi \rho_0}} \nabla {\bf B}_0 \right) 
-\frac{1}{\rho_0} \nabla P - {\bf Z}_\mp \cdot \nabla
{\bf Z}_\pm , \label{eq:Zpm1}
\end{eqnarray}
where ${\bf V}_A \equiv {\bf B}_0 / \sqrt{4 \pi \rho_0}$ is the
Alfv\'{e}n velocity vector, ${\bf U}$ is the outflow velocity, and
$P$ is the perturbation in the total pressure \citep[also see][]
{Dmitruk2001, Verdini2007, Buchlin2007, Cranmer2007}. In the present
paper we neglect flows along the field lines, so we set ${\bf U} =
0$. The terms in equation (\ref{eq:Zpm1}) involving ${\bf V}_A$ and
${\bf B}_0$ can be simplified by using equations (\ref{eq:HB}) and
(\ref{eq:dB0hat}), which yields
\begin{eqnarray}
\frac{\partial {\bf Z}_\pm} {\partial t} & = &
\pm v_A \hat{\bf B}_0 \cdot \nabla {\bf Z}_\pm 
\mp \frac{B_0}{2} \frac{d}{dz} \left( \frac{1}{\sqrt{4 \pi \rho_0}}
\right) \left( {\bf Z}_\pm  - {\bf Z}_\mp \right) \nonumber \\  
 &  & \pm \frac{1}{2} \frac{1}{\sqrt{4 \pi \rho_0}} \frac{dB_0}{dz}
{\bf Z}_\mp -\frac{1}{\rho_0} \nabla P - {\bf Z}_\mp \cdot \nabla
{\bf Z}_\pm . \label{eq:Zpm2}
\end{eqnarray}
We now take the curl of the above equation and evaluate the component
parallel to the mean magnetic field. Using equation (\ref{eq:dB0hat})
to evaluate the derivatives of $\hat{\bf B}_0$, we find
\begin{eqnarray}
\frac{\partial \omega_\pm} {\partial t} & = &
\pm v_A \left( \hat{\bf B}_0 \cdot \nabla \omega_\pm -
\frac{\omega_\pm}{2 H_B} \right) 
\mp \frac{B_0}{2} \frac{d}{dz} \left( \frac{1}{\sqrt{4 \pi \rho_0}}
\right) \left( \omega_\pm  - \omega_\mp \right) \nonumber \\  
 &  & \pm \frac{1}{2} \frac{1}{\sqrt{4 \pi \rho_0}} \frac{dB_0}{dz}
\omega_\mp - {\bf Z}_\mp \cdot \nabla \omega_\pm \pm {\cal N} ,
\label{eq:Zpm3}
\end{eqnarray}
where $\omega_\pm \equiv \hat{\bf B}_0 \cdot ( \nabla \times
{\bf Z}_\pm )$ are the vorticities associated with the Elsasser
variables, and ${\cal N}$ is given by
\begin{equation}
{\cal N} =
- \frac{\partial Z_{-,x}}{\partial x}
  \frac{\partial Z_{+,y}}{\partial x}
- \frac{\partial Z_{-,y}}{\partial x}
  \frac{\partial Z_{+,y}}{\partial y}
+ \frac{\partial Z_{-,x}}{\partial y}
  \frac{\partial Z_{+,x}}{\partial x}
+ \frac{\partial Z_{-,y}}{\partial y}
  \frac{\partial Z_{+,x}}{\partial y} .
\end{equation}
Combining the linear terms in equation (\ref{eq:Zpm3}) and using
$(\omega_+ - \omega_-)/2 = v_A \alpha$, we find that the above
equations are identical to equations (\ref{eq:dodt_pm}) and
(\ref{eq:N}). Therefore, the present formalism is equivalent to that
used in previous work based on Elsasser variables.

\section{B. Numerical Methods}
\label{sect:Num}

The four quantities $\omega$, $\alpha$, $h$ and $f$ are determined
by equations (\ref{eq:alpha}), (\ref{eq:omega}), (\ref{eq:dodt}) and
(\ref{eq:dhdt}) derived in section \ref{sect:RMHD}. In this Appendix
we describe the numerical methods used for solving these equations.
Let $r$ be the radial distance to the tube axis ($r < R(z)$),
$\varphi$ the azimuth angle, and $z$ the distance along the tube.
It is convenient to write the stream function as $f (\xi,\varphi,
z,t)$, where $\xi \equiv r/R$ is the {\it fractional} distance, and
similar for $h$, $\omega$ and $\alpha$. Then spatial derivatives along
the background field can be written as partial derivatives at constant
$\xi$; for example, $\hat{\bf B}_0 \cdot \nabla f = \partial f
/ \partial z$ in equation (\ref{eq:dhdt}). In this way the vector
$\hat{\bf B}_0$ can be eliminated from the equations.

\begin{figure*}
\epsscale{0.77}
\plotone{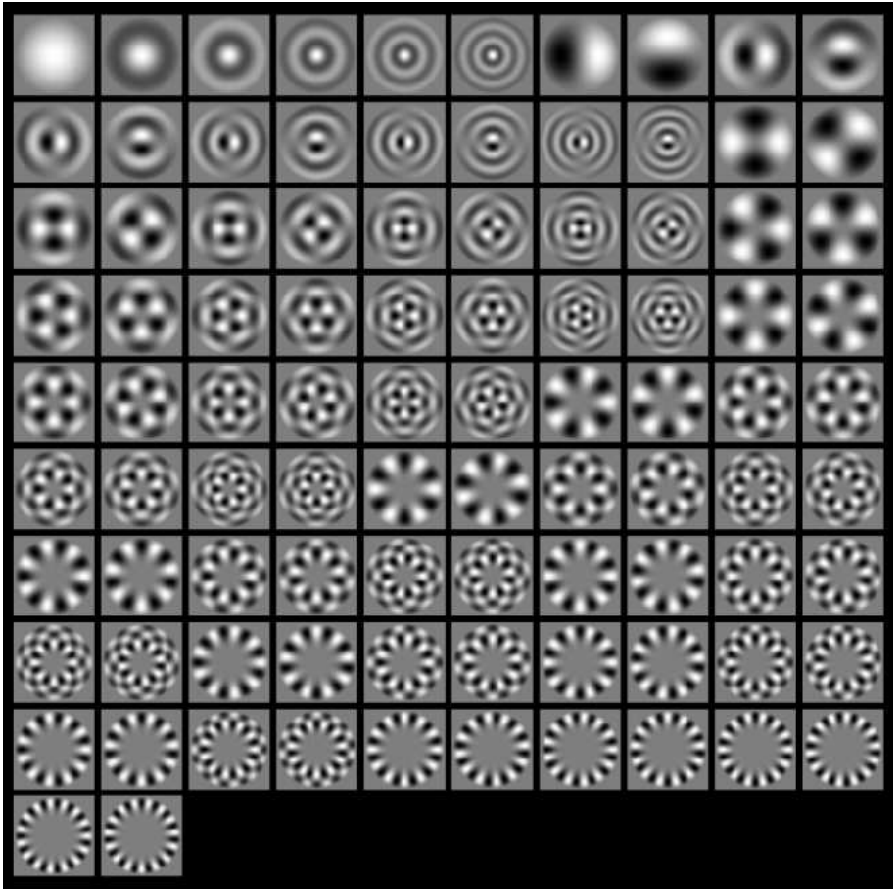}
\caption{The 92 basis functions $F_i (x,y)$ for $a_{max} = 20$. The
two driver modes are shown as the seventh and eighth image of the top
row.}
\label{fig:func2}
\end{figure*}

The transverse structure of the waves is described using a spectral
method. We use a finite set of basis functions $F_i (\xi,\varphi)$,
where index $i$ enumerates the basis functions ($i = 1, \cdots, N$).
The functions are assumed to vanish at the tube wall, $F_i (1,\varphi)
= 0$, consistent with our requirement that ${\bf v}$ and ${\bf B}$ are
parallel to the wall. The basis functions are mutually orthogonal and
are normalized such that
\begin{equation}
\frac{1}{\pi} \int_{0}^{2 \pi} \int_0^1 F_i (\xi,\varphi)
F_j (\xi,\varphi) ~ \xi d \xi d \varphi = \delta_{ij} ,
\label{eq:ortho}
\end{equation}
where $\delta_{ij}$ is the Kronecker delta symbol. An arbitrary
spatial distribution $f(\xi,\varphi)$ can always be {\it projected}
onto the basis functions by evaluating
\begin{equation}
f_{\rm proj} (\xi,\varphi) = \sum_{i=1}^{N} f_i F_i (\xi,\varphi) ,
\end{equation}
where $f_i$ are the mode amplitudes:
\begin{equation}
f_i = \frac{1}{\pi} \int_{0}^{2 \pi} \int_0^1 f(\xi,\varphi)
F_i (\xi,\varphi) ~ \xi d \xi d \varphi .  \label{eq:fj}
\end{equation}
In the following we assume $f \approx f_{\rm proj}$, i.e., we suppress
high spatial frequencies that are not present in the basis functions.
The basis functions $F_i (\xi,\varphi)$ are chosen to be the
eigenmodes of the $\nabla_\perp^2$ operator:
\begin{equation}
R^2 \nabla_\perp^2 F_i = \lambda_i F_i ,
\end{equation}
where $\lambda_i$ is the dimensionless eigenvalue. The solutions
$F_i (\xi,\varphi)$ are proportional to $J_m (a_i \xi)$, where
$m = m_i$ is the azimuthal mode number ($m \ge 0$), $J_m(x)$ is the
Bessel function of the first kind, and $a_i$ is a zero of the Bessel
function, $J_m (a_i) \equiv 0$. The modes are ordered according to
their $m$-values (first the $m=0$ modes, then $m=1$ modes, etc.),
and for fixed $m$ they are ordered according to their $a$-values.
The non-axisymmetric modes always come in pairs, one proportional to
$\cos (m \varphi)$, the other proportional to $\sin (m \varphi)$,
which we count as two separate modes $i$ and $i+1$, respectively.
The eigenvalues of the modes are $\lambda_i = - a_i^2$. We find
all modes with $a_i$ below a prescribed maximum value, $a_{max}$.
For example, for $a_{max} = 20$ we find $N=92$ modes (6 modes with
$m=0$, 12 modes with $m=1$, 10 modes with $m=2$, etc., up to $m=15$).
The basis functions are shown in Figure \ref{fig:func2}.

As shown in section \ref{sect:RMHD}, all nonlinear terms in the
RMHD equations involve the bracket operator, defined in equation
(\ref{eq:bracket}). In polar coordinates this quantity can be written
as
\begin{equation}
b \equiv [f,g] = \frac{1}{r} \left(
\frac{\partial f} {\partial r} \frac{\partial g} {\partial \varphi} -
\frac{\partial f} {\partial \varphi} \frac{\partial g} {\partial r}
\right) ,
\end{equation}
where $f(r,\varphi)$ and $g(r,\varphi)$ are arbitrary functions.
Expanding $f$, $g$ and $b$ in terms of basis functions, we obtain
\begin{equation}
b_k = \frac{1}{R^2} \sum_{j=1}^{N} \sum_{i=1}^{N} M_{kji} f_j g_i ,
\label{eq:bk}
\end{equation}
where $M_{kji}$ is the coupling matrix:
\begin{equation}
M_{kji} \equiv \frac{1}{\pi} \int_{0}^{2 \pi} \int_0^1 F_k \left(
\frac{\partial F_j} {\partial \xi} \frac{\partial F_i}
{\partial \varphi} - \frac{\partial F_j} {\partial \varphi}
\frac{\partial F_i} {\partial \xi} \right) ~ d \xi d \varphi .
\end{equation}
Note that $M_{kji}$ is anti-symmetric with respect to $i$ and $j$,
and using partial integration one can show that $M_{kji} = M_{ikj}$,
where we use $\partial F_i / \partial \varphi = 0$ for $\xi = 0$.
It follows that the six matrix elements that couple any three modes
$i$, $j$ and $k$ are closely related:
\begin{equation}
M_{kji} = M_{ikj} = M_{jik} = - M_{kij} = - M_{ijk} = - M_{jki} .
\label{eq:M6}
\end{equation}
To compute these elements we only need to consider the case
$k > j > i$ (the ``unique'' elements). The three modes must have
compatible $\cos (m \varphi)$ or $\sin (m \varphi)$ dependence, and
for $m_i > 0$ we only need to consider the case $m_k = m_j + m_i$.
We find that for $a_{max} = 20$ there are 7662 unique matrix
elements.

The quantities $f$, $h$, $\omega$ and $\alpha$ can be written in
terms of basis functions, for example:
\begin{equation}
f(\xi,\varphi,z,t) = \sum_{i=1}^{N} f_i (z,t) F_i(\xi,\varphi), 
\label{eq:phi}
\end{equation}
where $f_i (z,t)$ are the amplitudes of the different transverse
modes. Inserting these expressions into equations (\ref{eq:dodt}) and
(\ref{eq:dhdt}), and using expression (\ref{eq:bk}), we find
\begin{eqnarray}
\frac{\partial \omega_k} {\partial t} & = &
v_A^2 \frac{\partial \alpha_k} {\partial z} + 
\frac{1}{R^2} \sum_{j=1}^{N} \sum_{i=1}^{N} M_{kji}
( v_A^2 \alpha_j h_i - \omega_j f_i ) - \nu_k \omega_k ,
\label{eq:dodt_k} \\
\frac{\partial h_k} {\partial t} & = & 
\frac{\partial f_k} {\partial z} + \frac{f_k} {H_B} +
\frac{1}{R^2} \sum_{j=1}^{N} \sum_{i=1}^{N} M_{kji} f_j h_i 
- \nu_k h_k ,
\label{eq:dhdt_k}
\end{eqnarray}
where 
\begin{equation}
\alpha_k = (a_k/R)^2 h_k, ~~~~~ \mbox{and} ~~~~~
\omega_k = (a_k/R)^2 f_k , \label{eq:deriv2}
\end{equation}
and we added artificial damping terms. The damping rate $\nu_k$ is
assumed to be independent of time $t$ and position $z$ along the tube,
but depends on the perpendicular wavenumber $a_k$ to the sixth power
(``hyperdiffusion''):
\begin{equation}
\nu_k = \nu_{max} \left( \frac{a_k} {a_{max}} \right)^6 ,
\label{eq:nuk}
\end{equation}
where $\nu_{max}$ is the damping rate of the highest wavenumber modes
(we use $\nu_{max} = 0.7$ $\rm s^{-1}$). The alternative form of the
dynamical equations, given by equation (\ref{eq:dodt_pm}), yields
\begin{equation}
\frac{\partial \omega_{\pm,k}} {\partial t} = \pm v_A
\frac{\partial \omega_{\pm,k}} {\partial z} - v_A\frac{d v_A}
{dz} \alpha_k + \cdots . \label{eq:dodt_pmk}
\end{equation}
Here we omit the nonlinear and damping terms because those terms will
be evaluated directly from equations (\ref{eq:dodt_k}) and
(\ref{eq:dhdt_k}) (see below).

In order to accurately describe the Alfv\'{e}n wave propagation, we use
a grid of positions $z_n$ such that the Alfv\'{e}n travel time $\Delta
\tau$ between neighboring grid points is exactly equal to the time
step $\Delta t_0$ of the calculation:
\begin{equation}
\frac{z_{n+1} - z_n} {v_{A,n+1/2}} = \Delta t_0 = \mbox{constant} .
\end{equation}
Here $v_{A,n+1/2}$ is the average speed between grid points $n$ and
$n+1$. The Alfv\'{e}n travel time from the TR to any position $z$ in the
corona is
\begin{equation}
\tau (z) - \tau (z_{TR}) = \frac{L_{cor}} {v_A (z_{TR})}
\tilde{f} [u(z)] ,
\end{equation}
where $u(z) \equiv -1 + 2(z-z_{TR})/L_{cor}$, and $\tilde{f}(u)$ is
defined by
\begin{equation}
\tilde{f} (u) = \onehalf \int_{-1}^{u} \left( \frac{1-0.8 u^2}
{0.2} \right)^{-1/7} [ 1 + (\Gamma -1) (1 - u^2) ] ~ du .
\end{equation}
Here we used equations (\ref{eq:rho0}), (\ref{eq:B0}) and
(\ref{eq:tau}), and $v_A (z_{TR})$ is the Alfv\'{e}n speed at the coronal
base, $v_A (z_{TR}) \equiv B_{cor} / \sqrt{4 \pi \rho_0 (z_{TR})}$.
The grid of $u_n$ values is found by inverting the function
$\tilde{f} (u)$. To ensure that the total Alfv\'{e}n travel time
through the corona is a multiple of $\Delta t_0$, we make a slight
adjustment to the assumed coronal loop length $L_{cor}$.

We now describe the procedure for solving equations (\ref{eq:dodt_k})
and (\ref{eq:dhdt_k}). Let $\omega_{k,n} (t)$ be the vorticity for
mode $k$ at grid point $n$ and time $t$, and similar for the magnetic
flux function $h_{k,n} (t)$. To advance these quantities from time $t$
to $t^\prime = t + \Delta t_0$, we use operator splitting, i.e.,
each time step $\Delta t_0$ consists of two parts. In Part 1 we
compute the effects of {\it nonlinear mode coupling and damping},
and in Part 2 we determine the effects of {\it wave propagation and
reflection}. For Part 1 we can omit the propagation and reflection
terms, i.e., the first term on the right-hand side of equation
(\ref{eq:dodt_k}) and the first and second terms in equation
(\ref{eq:dhdt_k}). The resulting equations no longer contain any
$z$-derivatives, so the equations can be solved separately for each
position $n$ along the loop. For Part 1, advancing $\omega_{k,n}$ and
$h_{k,n}$ from time $t$ to $t^\prime$ is done using a fifth-order
Runge-Kutta method \citep[][] {Press1992}. The method uses adaptive
stepsize control, i.e., the time interval $\Delta t_0$ may be broken
up into two or more smaller intervals with $\Delta t < \Delta t_0$.
On the other hand, for Part 2 (wave propagation and reflection) we can
omit the nonlinear terms in the equations, so we instead use the
linearized form of equation (\ref{eq:dodt_pmk}), where the
Elsasser-like variables $\omega_{\pm,k,n}$ are computed from
equation (\ref{eq:pm}). The solution of equation (\ref{eq:dodt_pmk})
for time step $\Delta t_0$ can be written as 
\begin{equation}
\omega_{\pm,k,n} (t+\Delta t_0) = \omega_{\pm,k,n \pm 1} (t)
- \Delta t_0 \left( v_A \frac{dv_A}{dz} \right)_{n \pm 1/2}
\alpha_{k,n \pm 1/2} (t) , \label{eq:lin}
\end{equation}
where $\alpha_{k,n \pm 1/2}$ is the average value of $\alpha$ in
between neighboring grid points. The first term in equation
(\ref{eq:lin}) describes wave propagation, and the second term
describes the coupling due to gradients in Alfv\'{e}n speed
(gradients in the TRs are treated separately, see below). Note
that step 2 can be done separately for each mode $k$ because there is
no mode coupling in the linearized equations. Therefore, in each time
step $\Delta t_0$ the effects of both linear and nonlinear terms in
the equations are taken into account.

In our model there are two chromosphere-corona TRs, one at each end of
the coronal loop. At these TRs the mass density (and therefore the
Alfv\'{e}n speed) changes discontinuously with position, which causes
strong wave reflection. Let $z_{TR}$ be the position of one of these
TRs. In our model there are two grid points associated with this
position: $z_1$ just below the discontinuity, and $z_2$ just above
it. Let $v_{A,1}$ and $v_{A,2}$ be the Alfv\'{e}n speeds at these points,
and let $f_{1,\pm}$ and $f_{2,\pm}$ be the corresponding inward
and outward propagating waves (note that $v_{A,1} < v_{A,2}$ for the
first TR, and $v_{A,1} > v_{A,2}$ for the second one). The waves
$f_{1,+}$ and $f_{2,-}$ that propagate away from the TR are
assumed to be linear combinations of the waves $f_{1,-}$ and
$f_{2,+}$ that propagate towards it:
\begin{eqnarray}
f_{1,+} & = & c_{11} f_{1,-} + c_{12} f_{2,+} , \\
f_{2,-} & = & c_{21} f_{1,-} + c_{22} f_{2,+} ,
\end{eqnarray}
where the $c_{ij}$'s are reflection and transmission coefficients.
Then the stream functions in the two regions are given by
\begin{eqnarray}
f_1 & = & \onehalf [(1+c_{11}) f_{1,-} + c_{12} f_{2,+}] , \\
f_2 & = & \onehalf [c_{21} f_{1,-} + (1+c_{22}) f_{2,+}] , \\
h_1 & = & \onehalf [-(1-c_{11}) f_{1,-} + c_{12} f_{2,+}] /v_{A,1} ,\\
h_2 & = & \onehalf [-c_{21} f_{1,-} + (1-c_{22}) f_{2,+}] /v_{A,2} .
\end{eqnarray}
At $z_{TR}$ both the velocity and the magnetic field perturbations must
be continuous \citep[see][]{Chandrasekhar1961}, so we require $f_1
= f_2$ and $h_1 = h_2$. Moreover, these conditions must be
satisfied for any value of $f_{1,-}$ and $f_{2,+}$. It follows
that the coefficients $c_{ij}$ are given by
\begin{equation}
c_{11} = - c_{22} = \frac{c-1}{c+1}, ~~~~
c_{12} = \frac{2}{c+1}, ~~~~ c_{21} = \frac{2 c}{c+1} ,
\end{equation}
where $c \equiv v_{A,2} / v_{A,1}$ is the ratio of Alfv\'{e}n speeds.

The code was tested in various ways. We studied the propagation of
Alfv\'{e}n wave packets in a uniform flux tube to verify that no wave
dispersion occurs. We also considered nonuniform tubes and verified
that energy is conserved in wave reflections. For the stratified loop
we verified that the total wave energy (integrated over the entire
volume of the tube) is constant in time when a wave packet is present
in the initial conditions, the footpoints are held fixed, and the
damping rate $\nu_0 = 0$. This demonstrates that wave energy is
conserved by the nonlinear interactions. For the full model we
verified that the energy injected by footpoint motion minus the energy
dissipated by damping equals the rate of change of the total energy to
an accuracy of about 5\% of the net input rate.

\section{C. Energy Equation for Alfv\'{e}n Waves}
\label{sect:Energy}

We first derive expressions for the averages of vector quantities over
the cross section of the flux tube. The rms velocity $\Delta v_{rms}
(z,t)$ is given by
\begin{eqnarray}
( \Delta v_{rms})^2 & \equiv & < | {\bf v} |^2 >
= \frac{1}{\pi R^2} \int_0^{2\pi} \int_0^R | \nabla f |^2 ~ r dr
d \varphi 
= \frac{1}{\pi R^2} \int_0^{2\pi} \int_0^R f ~ \omega ~ r dr d \varphi
\nonumber \\
 & = & \frac{1} {\pi} \int_0^{2\pi} \int_0^1 \left( \sum_{j=1}^N
f_j F_j \right) \left( \frac{1}{R^2} \sum_{i=1}^N a_i^2 f_i F_i
\right) ~ \xi d \xi d \varphi
= \frac{1} {R^2} \sum_{k=1}^{N} a_k^2 f_k^2 , \label{eq:DVrms}
\end{eqnarray}
where $< \cdots >$ denotes the spatial average over the cross section
of the tube. Here we used partial integration to obtain $\omega$
($=-\nabla_\perp^2 f$), and we used equations (\ref{eq:ortho}),
(\ref{eq:phi}) and (\ref{eq:deriv2}). Similarly, the rms magnetic
fluctuation $\Delta B_{rms} (z,t)$ is given by
\begin{equation}
( \Delta B_{rms} )^2 \equiv < | {\bf B}_\perp |^2 >
= \frac{B_0^2}{\pi R^2} \int_0^{2\pi} \int_0^R | \nabla h |^2 ~ r dr
d \varphi = \frac{B_0^2} {R^2} \sum_{k=1}^{N} a_k^2 h_k^2 .
\label{eq:DBrms} 
\end{equation}
and the rms values of the Elsasser variables are
\begin{equation}
Z_\pm^2 \equiv < | {\bf Z}_\pm |^2 > = \frac{1} {R^2}
\sum_{k=1}^{N} a_k^2 (f_k \pm v_A h_k )^2 . \label{eq:DZrms}
\end{equation}
It follows from equations (\ref{eq:DVrms}) and (\ref{eq:DBrms}) that
the contributions of mode $k$ to the velocity and magnetic power
spectra are
\begin{equation}
P_{V,k} (z,t) \equiv (a_k/R)^2 f_k^2 ~~~ \mbox{and} ~~~~
P_{B,k} (z,t) \equiv B_0^2 (a_k/R)^2 h_k^2 .
\label{eq:Power}
\end{equation}
We now derive the energy equation in the context of the present model.
The kinetic and magnetic energy densities are given by
\begin{eqnarray}
E_{kin} (z,t) & = & \onehalf \rho_0 ( \Delta v_{rms} )^2 =
\frac{\rho_0}{2R^2} \sum_{k=1}^{N} a_k^2 f_k^2 , \label{eq:ekin} \\
E_{mag} (z,t) & = & \frac{( \Delta B_{rms} )^2} {8 \pi} = \frac{B_0^2}
{8\pi R^2} \sum_{k=1}^{N} a_k^2 h_k^2 . \label{eq:emag} 
\end{eqnarray}
Multiplying equation (\ref{eq:dodt_k}) by $\rho_0 f_k$ and summing
over $k$, we obtain the following equation for the kinetic energy
density:
\begin{equation}
\frac{\partial E_{kin}} {\partial t} = \sum_{k=1}^{N} f_k \left[
a_k^2 \frac{B_0^2} {4 \pi} \frac{\partial} {\partial z} \left(
\frac{h_k} {R^2} \right) + \frac{1}{R^4} \sum_{j=1}^{N} \sum_{i=1}^{N}
M_{kji} a_j^2 \left( \frac{B_0^2}{4\pi} h_j h_i - \rho_0 f_j f_i
\right) \right] - Q_{kin} , \label{eq:ekin_dt}
\end{equation}
where $Q_{kin}$ is the rate of kinetic energy loss due to damping:
\begin{equation}
Q_{kin} (z,t) \equiv \frac{\rho_0} {R^2} \sum_{k=1}^{N} \nu_k a_k^2
f_k^2 . \label{eq:qkin}
\end{equation}
Similarly, multiplying equation (\ref{eq:dhdt_k}) by $B_0^2/(4 \pi)
(a_k/R)^2 h_k$ yields an equation for the magnetic energy density:
\begin{equation}
\frac{\partial E_{mag}} {\partial t} = \sum_{k=1}^{N} a_k^2 \frac{h_k}
{R^2} \frac{B_0^2}{4 \pi} \left[ \frac{\partial f_k}
{\partial z} + \frac{1}{B_0} \frac{dB_0} {dz} f_k +
\frac{1}{R^2} \sum_{j=1}^{N} \sum_{i=1}^{N} M_{kji} f_j h_i \right]
- Q_{mag} , \label{eq:emag_dt}
\end{equation}
where $Q_{mag}$ is the rate of magnetic energy loss:
\begin{equation}
Q_{mag} (z,t) \equiv \frac{B_0} {4 \pi R^2} \sum_{k=1}^{N} \nu_k a_k^2
h_k^2 . \label{eq:qmag}
\end{equation}
Adding equations (\ref{eq:ekin_dt}) and (\ref{eq:emag_dt}) yields
\begin{equation}
\frac{\partial E} {\partial t} + B_0 \frac{\partial} {\partial z}
\left( \frac{F_A} {B_0} \right) = - Q , \label{eq:etot_dt}
\end{equation}
where $E(z,t) \equiv E_{kin} + E_{mag}$ is the total energy density,
$Q(z,t) \equiv Q_{kin} + Q_{mag}$ is the total dissipation rate,
$B_0 (z)$ represent the diverging geometry of the background field,
and $F_A (z,t)$ is the Alfv\'{e}n wave energy flux averaged over the
cross-section of the flux tube:
\begin{equation}
F_A (z,t) \equiv - \frac{B_0^2} {4 \pi R^2} \sum_{k=1}^{N} a_k^2
f_k h_k . \label{eq:flux}
\end{equation}
Note that the terms involving $M_{kji}$ drop out of the energy
equation (\ref{eq:etot_dt}). This is a result of equation
(\ref{eq:M6}), which shows that the matrix $M_{kji}$ is anti-symmetric
with respect to interchange of any two indices. Therefore, the
nonlinear terms in equations (\ref{eq:dodt_k}) and (\ref{eq:dhdt_k})
do not have a direct effect on energy transport along the tube, but
they are of course responsible for exciting the high wavenumber modes
that are subject to damping ($Q$). Therefore, our numerical scheme
using basis functions $F_i (\xi,\varphi)$ is excellent for studying
the energetics of the waves.

\end{document}